\documentclass[authoryear,final,1p,12pt]{elsarticle}

\usepackage{amssymb}
\usepackage{amsmath,amsfonts}
\usepackage{algorithmic}
\usepackage{array}
\usepackage{multirow}
\usepackage{textcomp}
\usepackage{stfloats}
\usepackage{url}
\usepackage{verbatim}
\usepackage{graphicx}
\usepackage{subfigure}
\usepackage{balance}
\usepackage[dvipsnames]{xcolor}
\usepackage{setspace}

\begin{document}

\begin{frontmatter}

\title{A Geometric Model with Stochastic Error for Abnormal Motion Detection of Portal Crane Bucket Grab}
\author[label1]{Baichen Yu}
\author[label2]{Xiao Wang\corref{cor}}
\author[label1]{Hansheng Wang}
\affiliation[label1]{organization={Guanghua School of Management, Peking University},
            addressline={No.5 Yiheyuan Road}, 
            city={Beijing},
            postcode={100871}, 
            country={China}}
\affiliation[label2]{organization={School of Mathematics and Statistics, Qingdao University},
            addressline={N0.308 Ningxia Road}, 
            city={Qingdao},
            postcode={266071}, 
            country={China}}
\cortext[cor]{Corresponding author, e-mail: xwang1020@qdu.edu.cn.}

\begin{abstract}
Abnormal swing angle detection of bucket grabs is crucial for efficient harbor operations. In this study, we develop a practically convenient swing angle detection method for crane operation, requiring only a single standard surveillance camera at the fly-jib head, without the need for sophisticated sensors or markers on the payload. Specifically, our algorithm takes the video images from the camera as input. Next, a fine-tuned `the fifth version of the You Only Look Once algorithm' (YOLOv5) model is used to automatically detect the position of the bucket grab on the image plane. Subsequently, a novel geometric model is constructed, which takes the pixel position of the bucket grab, the steel rope length provided by the Programmable Logic Controller system, and the optical lens information of the camera into consideration. The key parameters of this geometric model are statistically estimated by a novel iterative algorithm. Once the key parameters are estimated, the algorithm can automatically detect swing angles from video streams. Being analytically simple, the computation of our algorithm is fast, as it takes about 0.01 seconds to process one single image generated by the surveillance camera. Therefore, we are able to obtain an accurate and fast estimation of the swing angle of an operating crane in real-time applications. Simulation studies are conducted to validate the model and algorithm. Real video examples from Qingdao Seaport under various weather conditions are analyzed to demonstrate its practical performance.\\


\end{abstract}

\end{frontmatter}

\section{Introduction}\label{intro}

Collisions caused by excessive swing of the payload of portal cranes, as one of the main concerns for safe operation in seaport transportation, could result in severe human casualties and substantial financial losses \citep{sadeghi2021systematic}. In this regard, an appropriate solution for real time estimation of the swing is needed for day-to-day operation in a seaport. This becomes the key motivation of this work and leads to unique contributions to the literature. The details of our motivation and contribution are separately presented below.

\subsection{Motivation}\label{motiv}

As a crucial component in the global economy, sea transportation is a highly complicated process with many critical stages, among which seaport transportation is arguably the most important one. Seaport transportation refers to the operation of transferring goods from the seaport to waiting cargo ships, or vice versa. This is a very heavy-duty process, which cannot be accomplished by manpower. Therefore, a large number of portal cranes have to be used. Consider for example Qingdao Seaport, which is one of the largest seaports in China for sea transportation and international trade. A total of 85 cranes have been used by Qingdao Seaport. For day-to-day crane operation, the arguably most critical concern is safety. As \cite{lam2007crane} reported in TOC ASIA, there were more than 12 accidents happened in the year of 2006, which caused millions of dollars in loss at Portek. The catastrophic outcomes of these accidents highlight the importance of maintaining, designing, and implementing strict and robust processes for safety assurance \citep{im2020crane}.

To gain some intuitive understanding, we provide in Figure \ref{Crane} a photo taken from Qingdao Seaport. Note that there is a steel wire rope for a standard portal crane, connecting the fly-jib head and the bucket grab; see Figure \ref{Crane} for an illustration of the portal crane. Due to the effect of gravity, the steel wire rope should be fully tightened. In an ideal situation, we should expect the steel wire rope to be fully aligned with the normal force of gravity. This is considered a safe situation. However, for an operating portal crane, its steel wire rope can never stay in a fully static position. The random swing of the steel wire rope is inevitable, and it could be even worse with bad weather. The swing leads to a non-zero angle between the steel wire rope and the normal force of gravity. For convenience, we refer to this angle as a swing angle. It could be extremely dangerous if the swing angle is too large. For safe operations, the swing angle should be controlled as small as possible. If a too large swing angle happens, an alarm signal should be generated. Accordingly, an appropriate anti-swing measure must be undertaken immediately.

In this regard, practitioners in Qingdao Seaport face two challenges. The first challenge is measuring. That is how to measure the swing angle accurately. The second challenge is monitoring. That is how to monitor the bucket grab working conditions (including the swing angle) carefully and continuously. The common practice in this regard is subjective eyeball inspection for both measuring and monitoring. This is obviously an inaccurate and unreliable solution. Therefore, we are motivated to develop an automatic solution with the desired accuracy and reliability. Our solution relies on a high-resolution surveillance camera placed on the fly-jib head that captures the bucket grab continuously. This leads to a streaming of high-resolution images; see Figure \ref{firstframe} for some illustrations. Meanwhile, the Programmable Logic Controller (PLC) system provides the length of the steel wire rope. These two pieces of information, together with the camera lens information, provide us with a unique opportunity to measure the swing angle accurately. This leads to a novel methodology to be presented subsequently.

\subsection{Contributions}\label{contri}

We attempt to make a very unique contribution to the existing literature on safety assurance for portal crane operation. Compared with the existing methods (both the sensor based and computer vision based), the most important feature of our method is the easiness for implementation. More specifically, implementing our method entails neither sophisticated or expensive sensors nor professional cameras. Instead, one single standard surveillance camera with the desired resolution level is all we need. Furthermore, since the camera is placed on the fly-jib head and shoots directly on the bucket grab, the resulting images are highly informative and significantly less affected by other noisy backgrounds due to (for example) another crane standing nearby. Theoretically, we also contribute to the literature by providing a novel geometric model, which integrates three pieces of information neatly. These three pieces of information are, respectively, the image information from the surveillance camera, the steel wire rope length information from the PLC system, and the optical information of the surveillance camera. With the geometric model, the relationship between the steel wire rope length and the swing angle can be analytically derived. This leads to an accurate and fast estimate of the swing angle. Our main contribution is the development of a computer-vision based marker-free method for bucket grab swing angle estimation. Compared with the existing methods, our method is: (1) working environment friendly, (2) marker free, and (3) highly accurate in swing angle estimation.

The rest of the article is organized as follows. The related work is reviewed in Section \ref{review}. Section \ref{datacp} describes the dataset and explains the procedure of data preparation. The main methodology is presented in Section \ref{method}. The article is concluded with a short discussion and future directions in Section \ref{conclude}. All technical details are provided in the Appendices.

\section{Related work}\label{review}

The current work is related to multiple scientific research fields. They are, respectively, IT-based sea transportation management, deep learning and machine learning methods, sensor based crane control system, and computer-vision based crane control system. The relevant literature of each research fields is to be reviewed subsequently.

\subsection{IT-based sea transportation management}\label{ITstm}

Our research is related to information technology (IT) based sea transportation management, where the key issue is how to utilize IT to improve the effectiveness, efficiency, and reliability of sea transportation \citep{chen2022evolutionary}. For example, \cite{dubrovsky2002genetic} proposed an IT-based algorithm to generate a stowage scheme that minimizes container movements and thus reduces handling time at berth. \cite{maki2011new} developed a generic algorithm to optimize fuel cost and path safety. \cite{ting2014particle} proposed a particle swarm optimization approach to solve the berth allocation problem. \cite{hottung2016biased} developed an algorithm for solving the container pre-scheduling problem. \cite{de2019hybridizing} developed a model to capture the complexities of the fuelling issues, such as the selection of fuelling ports, total fuelling amount at a port, and others. \cite{mogale2023designing} developed a mathematical model to minimize total costs encompassing transportation cost, pipeline and retailers inventory cost, fixed cost of cross-dock, and carbon emission costs. 

\subsection{Deep learning and machine learning methods}\label{deepm}
As a rising trend, various deep learning and machine learning models have been developed for transportation \citep{yin2021deep}, supply chains \citep{sharma2020systematic}, and logistics planning \citep{jahani2023data}. For example, \cite{melanccon2021machine} used XGBoost \citep{chen2016xgboost} to predict service failures so that supply chain reliability can be improved. \cite{qi2023practical} utilized a multiquantile recurrent neural network \citep{wen2017multi,fan2019multi} to predict inventory replenishment amounts to reduce inventory costs and improve turnover rates. 
\cite{mousapour2023hybrid} used a convolutional neural network with a bidirectional gated recurrent unit \citep{chakrabarty2017context} to predict fake signals from financial markets. \cite{yang2023traffic} applied the long short-term memory model \citep{hochreiter1997long} to predict the traffic speed in a large traffic network.   \cite{ozarik2024machine} adopted the LASSO method \citep{tibshirani1996regression} to predict the score in candidate delivery sequences. Recently, \cite{zhan2024hybrid} integrated deep learning models within a decision-making framework so that the low-carbon transportation systems can be accurately assessed.

Moreover, several metaheuristic optimization and deep learning methods in complex system optimization have been developed for image data analysis. \cite{abdollahzadeh2024puma} proposed the Puma Optimizer (PO) with phase change mechanisms, which improved performance across various optimization tasks. For reviews of the Moth-Flame and Manta Ray Foraging optimization methods, we refer to \cite{zamani2024critical} and \cite{gharehchopogh2024advances}. Regarding image data analysis, \cite{ozbay2023interpretablefusion} combined interpretability and feature fusion for MRI brain tumor detection. \cite{ozbay2023interpretablepap} enhanced cervical cancer detection by using hybrid dilated convolution and spatial attention.

\subsection{Sensor based crane control system}\label{sens}
The research objective of this work is related to safety assurance for crane operation. There exist two different approaches. The first approach is the sensor-based approach. For example, \cite{kim2001new} proposed an acceleration sensor based anti-swing control system. The swing angle can be estimated by data collected from the sensors equipped on the trolley. \cite{sano2010anti} proposed an anti-swing crane control system, which used two state observer sensors to correct the swing angle. \cite{matsunaga2018sound} proposed a deflection angle measurement method by using two microphones attached to the trolley. \cite{miranda2019family} designed a family of anti-swing motion controllers for 2D cranes by integrating translational motion and swing angle coordinates. Recently, \cite{helma2021inertial} proposed a swing angle control system, where the raw acceleration and angular rate data obtained from inertial measurement units are used. Additionally, \cite{miranda2021robust} presented a model-free control scheme, which included a proportional derivative controller, a disturbance observer, and a compensation term with a coupling function.

\subsection{Computer-vision based crane control system}\label{coms}
The merit of the sensor based anti-swing motion control approach lies in its accuracy. However, the drawback is that they are relatively expensive \citep{clark2020multi}. As a useful alternative, computer-vision based approaches are considerably more cost efficient. In this regard, \cite{matsuo2004nominal} developed a crane load swing motion measurement system. The system uses data collected from Video Tracker G220 to compute the swing angle. \cite{kawai2009anti} proposed an anti-swing system, which measures the swing motion by a charge-coupled device camera installed on the side of the trolley. \cite{jung2012advanced} developed a robust sensing system, containing a smart vision camera, an infrared rays pass filter lens, and two inertial measurement units. \cite{okubanjo2018vision} proposed a vision based crane control system, which uses webcams for image data collection and uses high-speed computers for image processing and crane control. 

\subsection{The limitation of the existing methods}\label{them}
Despite the practical usefulness of those existing computer-vision based approaches, none of them can be readily used to solve the problem from Qingdao Seaport due to the following reasons. First, some are tailored to cranes of different designs and physical structures. For example, the method of \cite{kawai2009anti} is based on a mass damper type anti-sway system, which is unfortunately not available at Qingdao Seaport. Second, some require that the camera to be installed at specific positions, which are very different from our case. For instance, \cite{okubanjo2018vision} required the high resolution camera to be installed under the trolley of a crane. However, at Qingdao Seaport, the camera is installed at the fly-jib head. Third, some require professional visual sensors for visual information collection. For example, \cite{matsuo2004nominal} utilized a video tracker to compute the swing angle. Moreover, one can have markers painted on the bucket grab, which seems to be an easier solution. However, in practice, markers sometimes tend to fade quickly, sometimes even before a single task is completed. In contrast, our proposed algorithm requires only a standard surveillance camera installed on the fly-jib head. Compared with these more complex approaches, our proposed algorithm is much easier to implement. Therefore, our method is working environment friendly. Additionally, our algorithm is a marker-free method and requires no markers to be painted on the bucket grab. Furthermore, our method produces swing angle estimates with excellent accuracy.

\section{Data collection and preparation }\label{datacp}

In this section, we describe in detail how the video data are collected, and how the steel wire rope length and pixel positions of the bucket grab are extracted.

\subsection{Video data collection}\label{datac}

\begin{figure}[!ht]
    \centering
    \includegraphics[width=0.75\textwidth]{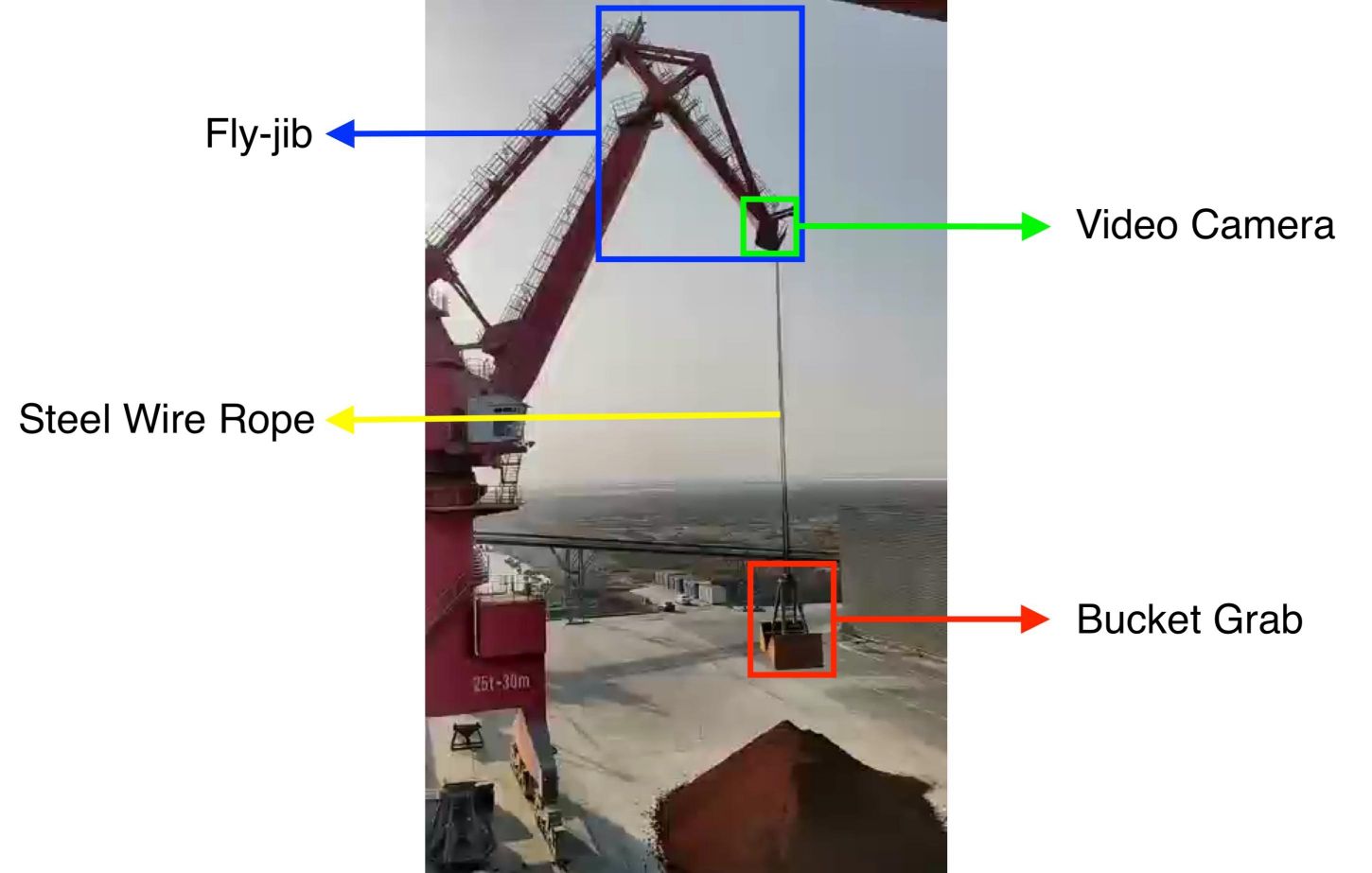}
    \caption{An operating crane with all important features labeled by different colors.}
    \label{Crane}
\end{figure}

\begin{figure}[!ht]
    \centering
    \includegraphics[width=0.6\textwidth]{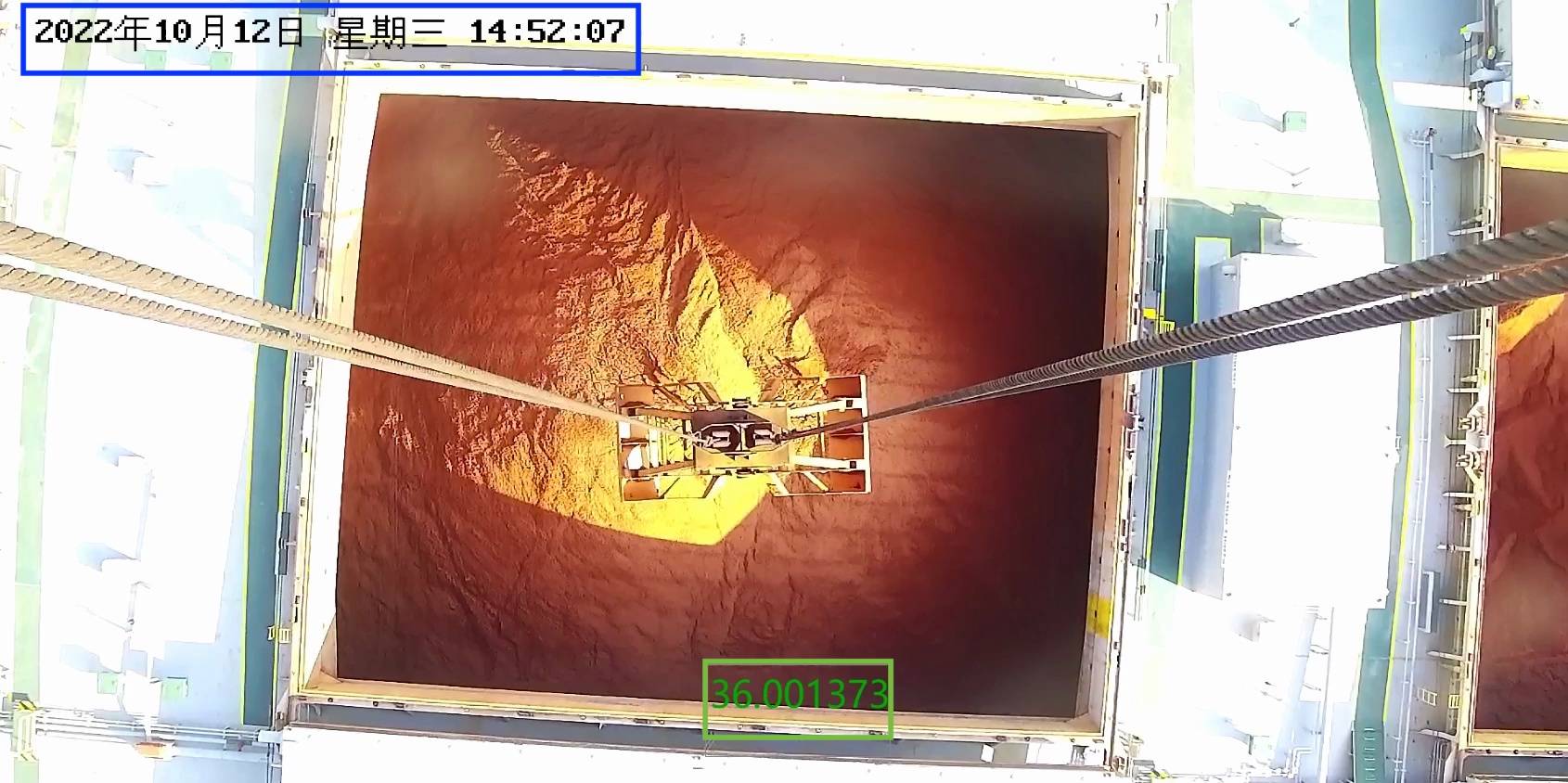}
    \caption{The graphical display of an image from a camera shooting to the bucket grab.}
    \label{firstframe}
\end{figure}

The image data are collected by a high-resolution surveillance video camera. The camera was installed on the fly-jib head of an operating crane located at Qingdao Seaport. To gain an intuitive understanding, we provide in Figure \ref{Crane} a real photo of an operating crane. In Figure \ref{Crane}, the large blue box frames the fly-jib, and the smaller green one frames the head of the fly-jib. The bucket grab framed in the red box is the main object captured by the video camera. Ideally, the placement angle of the camera should be carefully adjusted so that the bucket grab sits right in the center of the image; see Figure \ref{firstframe} for illustration. Lastly, there is a steel wire rope connecting the fly-jib and the bucket grab.

As mentioned before, the camera should be carefully installed so that the bucket grab in the video frame can be centered as much as possible; see Figure \ref{firstframe} for illustration. Once installed, the camera generates video files in MP4 format with $9$ frames recorded per second. This leads to a total of $3,699$ frames lasting over $6$ minutes. The date and time are also displayed in the image in Chinese; see the blue box in the top left corner of Figure \ref{firstframe}. In addition to that, the length of the steel wire ropes is  displayed in the green box of the video. As mentioned before, this is a piece of information provided by the PLC system. We next consider how to transfer the video data into structured data for subsequent analysis.

\subsection{Steel wire rope length extraction}\label{extractl}
\begin{figure}[!ht]
    \centering \includegraphics[width=0.7\textwidth]{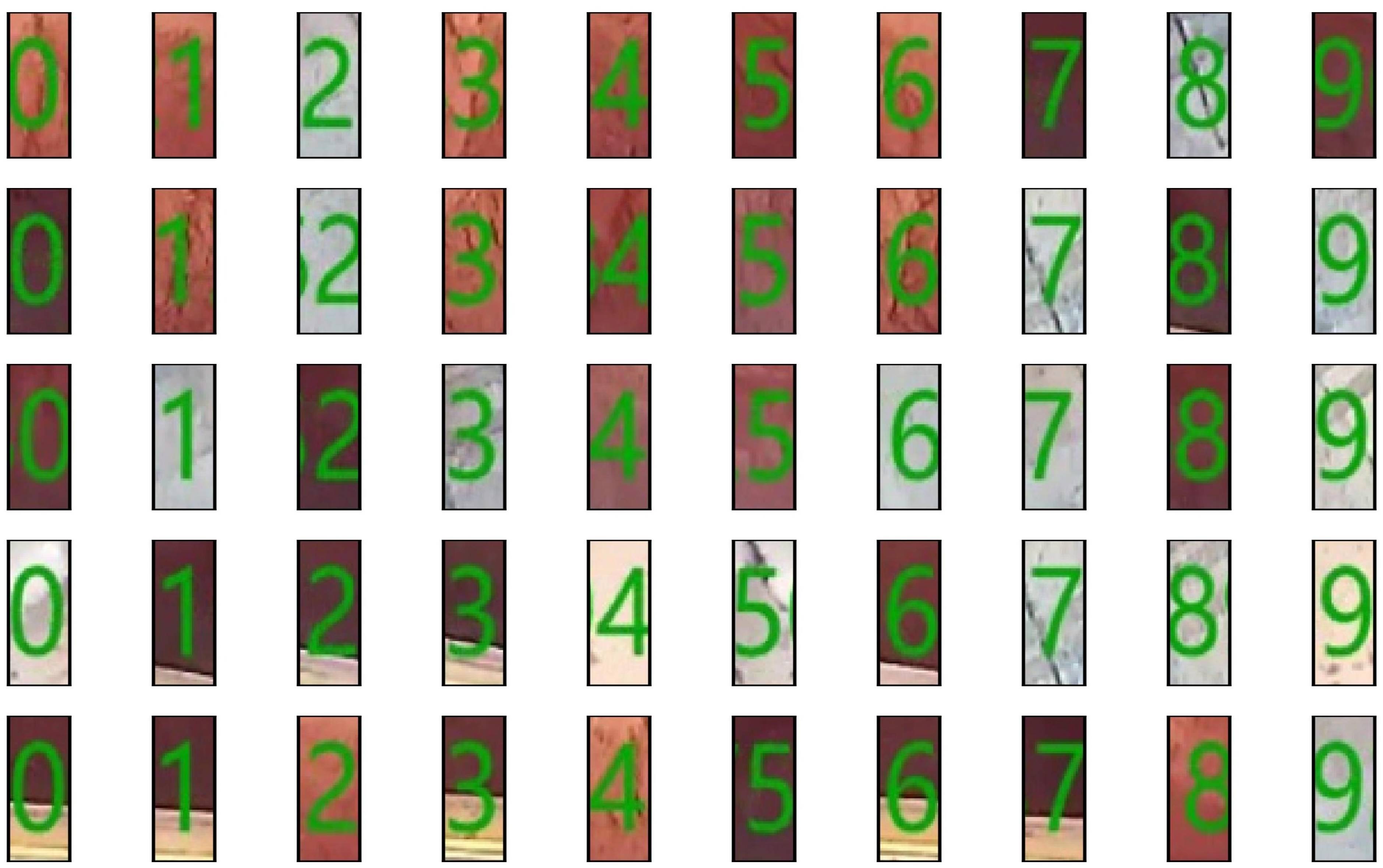}
    \caption{Illustration of the digital samples extracted from the images.}
    \label{cuttednum}
\end{figure}

The first piece of information needs to be extracted from the video frames is the length of the steel wire ropes. That is the digital information displayed in the green box in Figure \ref{firstframe}. Note that this information is displayed at a fixed pixel position in the image with a total of $8$ digits. Since their positions are fixed, we can easily extract the corresponding sub-images for every single digit as independent samples; see Figure \ref{cuttednum}. This leads to a total of $29,592$ sub-images, among which $320$ are manually labeled according to their digit categories. The labeled dataset is then randomly split into a training dataset and a testing dataset. The training dataset accounts for about $80\%$ of the labeled samples, while the rest are used for testing. Next, a classical LeNet-5 model of \cite{lecun1998gradient} can be trained on the training dataset. The resulting prediction accuracy is $100\%$ on the testing set. We then apply the model to all the images so that the length information of steel wire ropes at each time could be exactly extracted.

\subsection{Pixel positions of the bucket grab extraction}\label{extractbg}
\begin{figure*}[!ht]
 	\centering
 	\subfigure[The manually annotated bounding box]{\includegraphics[width=0.45\textwidth]{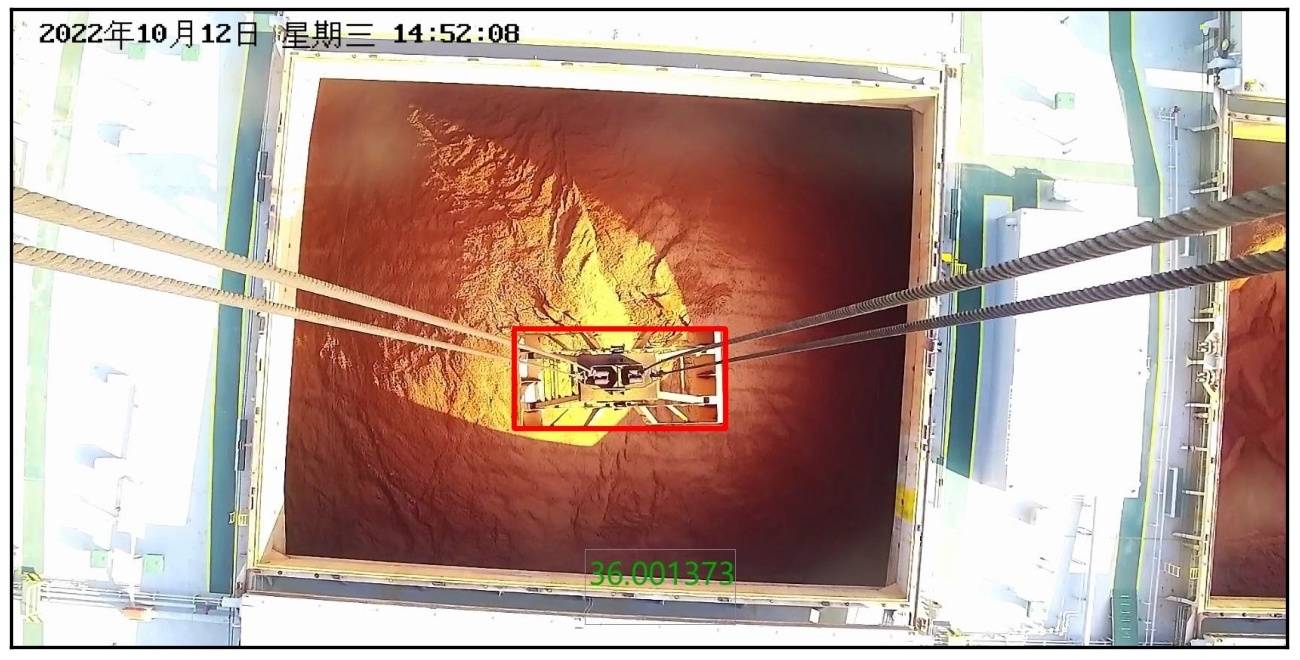}}
 	\hfil
 	\subfigure[The YOLOv5 generating bounding box]{\includegraphics[width=0.45\textwidth]{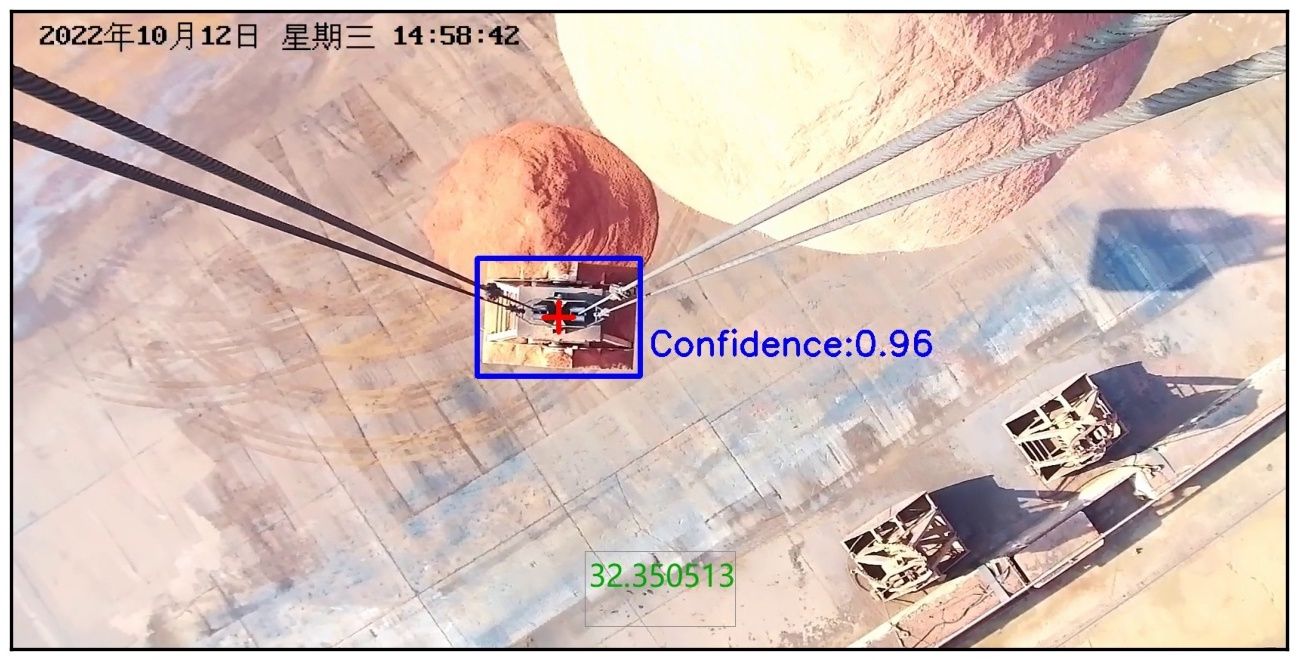}}
 	\caption{The illustration of manually annotated and YOLOv5 generated bounding boxes.}
 	\label{YOLOv5}
\end{figure*}

We next consider how to capture the pixel position of the bucket grab. To this end, we need to detect in the image the bucket grab automatically and accurately. Unlike the rope length digits, the pixel location of the bucket grab is random due to operation, weather conditions, or possibly other factors. To solve this problem, a total of $340$ images are used for model building. Those $340$ images are randomly extracted from an independent clip of continuous video record without any subjective selection. They are then manually annotated with the position of the bucket grab by tight bounding boxes. The position and the size of the bounding boxes are accurately recorded; see the left panel of Figure \ref{YOLOv5} for illustration. The annotated dataset is then randomly split into a training dataset and a validation dataset. The training dataset contains about $80\%$ of the annotated samples, while the rest are used for validation. A pre-trained `the fifth version of the You Only Look Once algorithm' (YOLOv5) model of \cite{glenn2022yolo} is fine-tuned on the training dataset, starting from the pre-trained model parameters trained on the Common Objects in Context (COCO) dataset \citep{lin2014microsoft}. The resulting model is then validated on the validation dataset with an mAP@50:95 metric (mean Average Precision among different Intersection Over Union from $50\%$ to $95\%$) as high as $0.948$. It seems that the accuracy level is already sufficient for addressing the real application we are currently dealing with, even though a larger sample size might yield better results.

With the help of the fine-tuned YOLOv5 model, a bounding box can be accurately generated for each image in the video frames. Then, the center of the YOLOv5 bounding box estimated by the YOLOv5 model (i.e., the red cross in the right panel of Figure \ref{YOLOv5}) is recorded as the center of the bucket grab. The pixel coordinate of this center is recorded and matched with the length of the steel wire rope for each image. Recall that we have a total of $3,699$ images. This leads to a dataset with $3,699$ records of the lengths of steel wire ropes and the pixel positions of bucket grab centers.

\section{Methodology}\label{method}

\begin{figure}[!ht]
    \centering
    \includegraphics[width=1\linewidth]{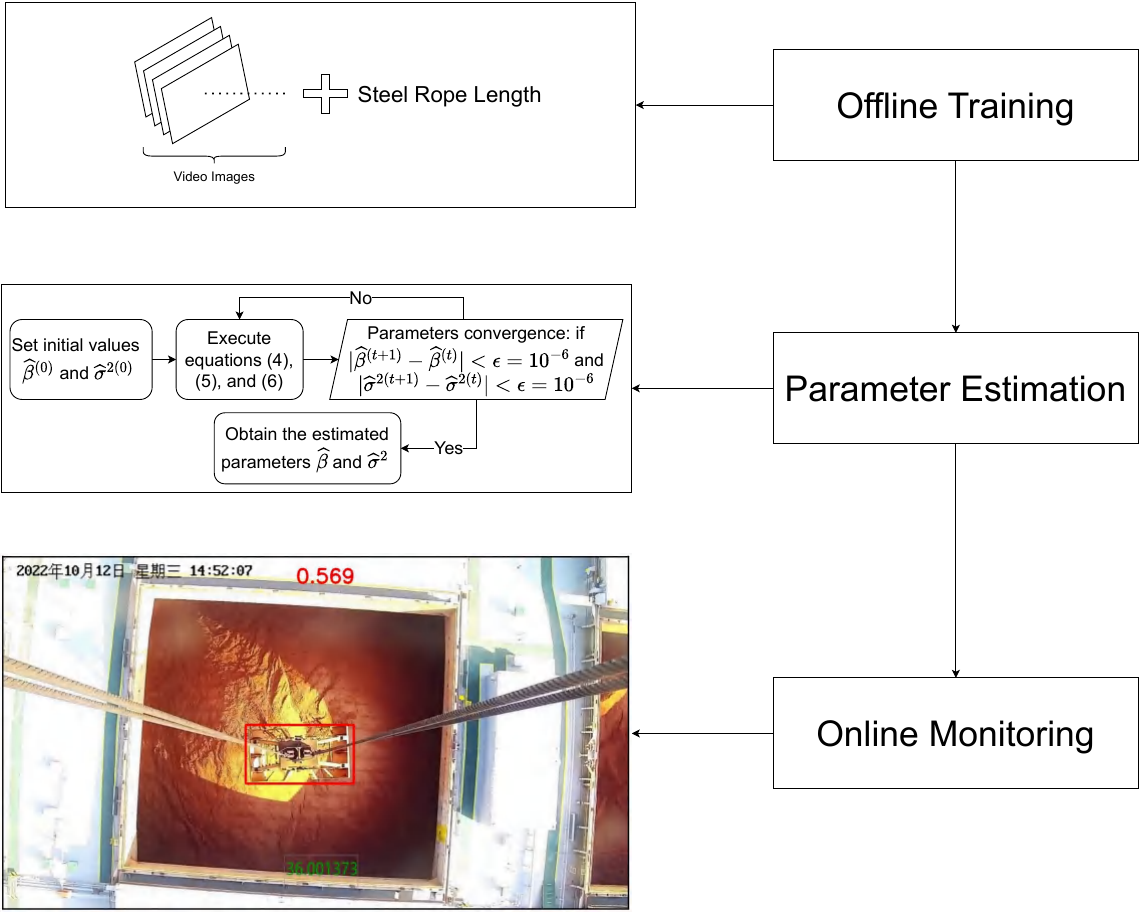}
    \caption{Main steps of the proposed algorithm. }
    \label{flowchart}
\end{figure}

We develop here a novel algorithm for swing angle estimation. The algorithm integrates information from both the surveillance camera (installed on the fly-jib head) and the steel wire rope length from the PLC system. A detailed flowchart is given in Figure \ref{flowchart}.

\subsection{A geometric model}\label{geometricmodel}

Recall that the swing angle is defined to be the angle formed by the steel wire rope and the direction of gravity. To estimate the swing angle, we only have two pieces of information in hand. One is the length of the steel wire rope and the other is the pixel position of the bucket grab center. Then how to estimate the angle with only these two pieces of information becomes the key issue. 

\begin{figure}[!ht]
    \centering
    \includegraphics[width=0.45\textwidth]{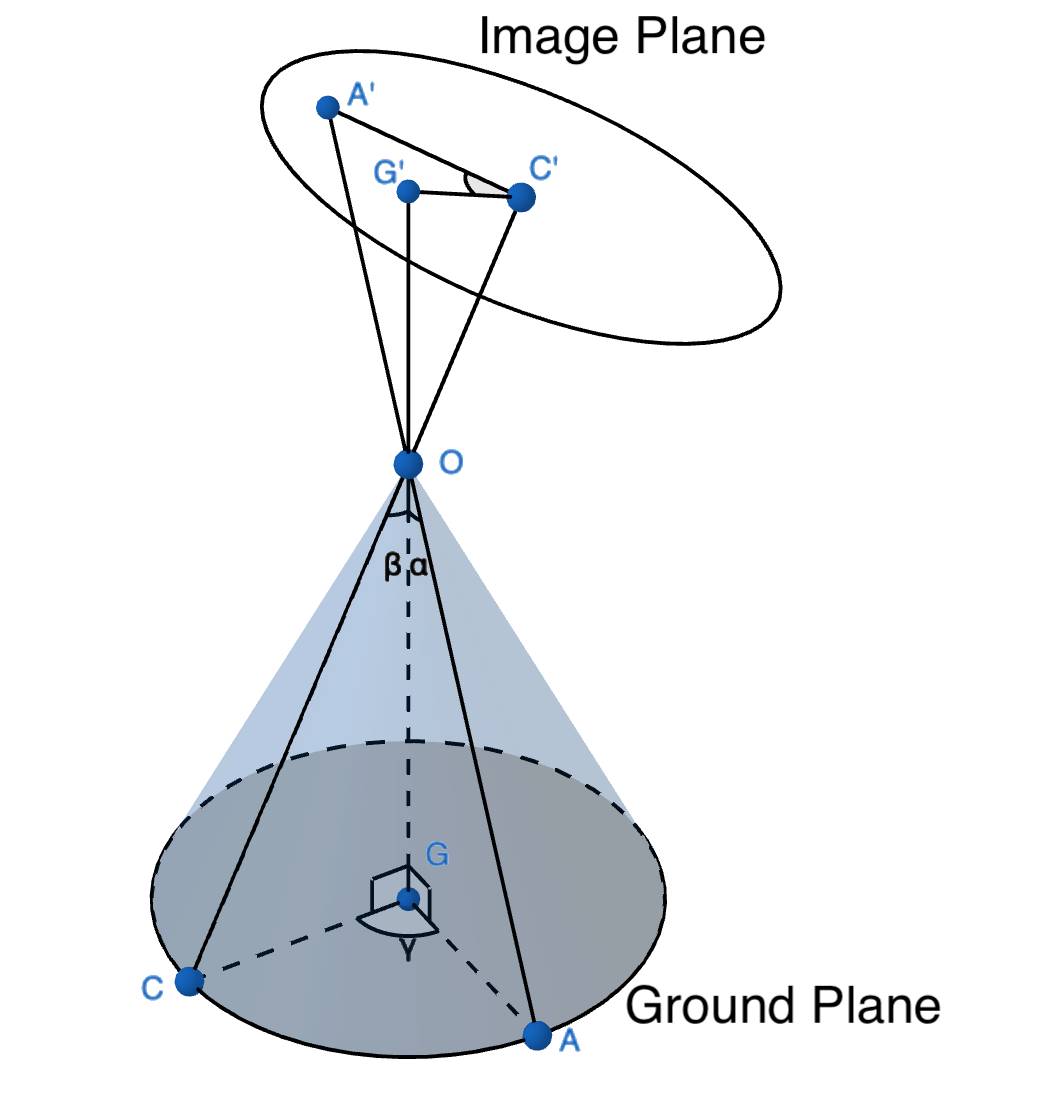}
    \caption{Illustration of the geometric model.}
    \label{geo_illus}
\end{figure}

To solve the problem, we develop here a geometric model as follows. The details are graphically displayed in Figure \ref{geo_illus}. Specifically, we use a point $O$ to represent the fly-jib head of the crane. That is the place to have one end of the steel wire rope connected. That is also the place having the video camera installed. Next, we use a vertical line $OG$ to represent the gravity direction. Therefore, $OG$ should be perpendicular to the ground plane. Then, we use line $OA$ to represent the steel wire rope connecting the fly-jib head and bucket grab. Here we use point $A$ to represent the center of the bucket grab. If there is no random perturbation, we should expect that the direction of both $OA$ and $OG$ to be perfectly overlapped with each other due to gravity. However, there exists a non-zero and random error $\alpha$ between $OA$ and $OG$ due to random swing. Note that $OA$ represents the length of the steel wire rope, which is known to us. Lastly, the point $G$ is particularly selected so that $OG$ is perpendicular to $GA$.

We next study the geometric properties of the video camera and the resulting image. Recall that the camera is installed on the fly-jib head, which is denoted by the point $O$. We use a straight line $OC$ to represent the principal axis of the camera. The point $C$ is selected so that $OG$ is perpendicular to $GC$. Ideally, we wish that the camera should be carefully installed so that $OC$ is also vertical to the ground plane as $OG$ does. However, non-negligible discrepancy seems inevitable in practice. Consequently, there also exists a non-zero but fixed angle $\beta$ between $OC$ and $OG$. Obviously, we should expect $\beta$ to be fairly small as long as the camera is installed carefully. Nevertheless, we can never assume it to be exactly $0$. A piece of empirical evidence supporting this claim is that the actually observed pixel positions of the bucket grabs are never perfectly overlapped with the center of the images. Instead, the probability distribution of the observed pixel positions for the bucket grab center seems obviously biased with respect to the image center. Once the camera is fixed, the line $OC$ is fixed so that the value of $\beta$ is also fixed without any further randomness. It is remarkable that in a $3$-dimensional Euclidean space with a randomly swing bucket grab, it is extremely hard to have the three points $G$, $A$, and $C$ perfectly aligned in one straight line. Consequently, another non-zero and random angle $\gamma$ should be formed between $GA$ and $GC$.

Lastly, we shall consider how the random position of the real bucket grab point $A$ to be projected onto the image. Here we use an ellipse contour on the top of Figure \ref{geo_illus} to represent a $2$-dimensional plane, which represents an image with infinite size. By definition, the image plane should be perpendicular to the principle axis of the video camera. Due to the fact that the camera's principle axis is not perfectly overlapped with the gravity direction as we explained in the previous paragraph, we cannot expect that the image plane and the ground plane to perfectly parallel with each other. We next extend the line $OA$ to intersect with the image plane at point $A'$. To accurately reflect its position, an appropriate coordinate system needs to be defined for the images. To this end, we extend $OC$ to intersect with the image plane at point $C'$. We then treat $C'$ as the origin of the image plane. Note that the length of $OC'$ reflects the focal length of the camera and is a fixed but unknown constant. Moreover, we extend $G$ to $G'$ on the image plane. Then we use $C'G'$ to represent the $0$-degree angle. Once the origin $C'$ and the $0$-degree reflection line $C'G'$ is given, the relative position of the projected bucket grab center on the image can be accurately described. Its analytical relationship with the swing angle can be derived subsequently. In addition, the minimal angle in our model is $0^{\circ}$, which happens when the crane is not operating and in a steady state. With reference to a technical protocol used by Shandong Port Group Co. Ltd (i.e. the company running the Qingdao Seaport), any swing angle larger than $10^{\circ}$ should be considered as danger. We therefore set the maximum angle of our algorithm to be $20^{\circ}$. 

\subsection{A geometric analysis}\label{geometricanalysis}

Note that the swing angle $\alpha$ is the key variable of interest, which is unfortunately not directly measurable. Our ultimate goal is to accurately recover the value of $\alpha$ by making use of the information from the image plane. To this end, two critical parameters are inevitably involved. The first parameter is the angle formed by the principle axis of the video camera and the gravity direction. For convenience, we refer to this as a camera angle and write it as $\beta$, which should be estimated by the information reflected on the image plane. The second parameter is the focal length $h=OC'$, which can be obtained subsequently in Section \ref{focallength}. Then, the sophisticated relationship needs to be analytically derived for the swing angle $\alpha$, the camera angle $\beta$, the random angle $\gamma$ formed by $GA$ and $GC$, the focal length $h$, and the observed distance on the image plane $m=A'C'$.

\begin{figure}[!ht]
    \centering
    \includegraphics[width=0.35\textwidth]{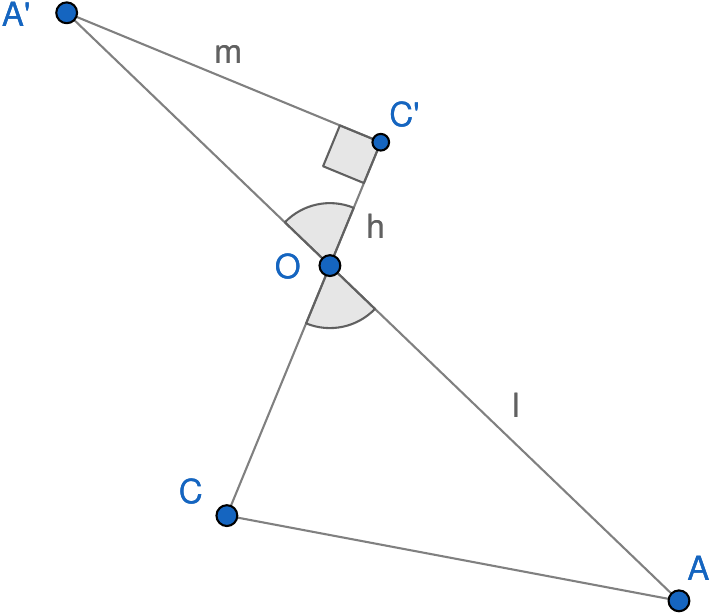}
    \caption{The extracted 2-dimensional plane formed by two intersected straight lines $AA'$ and $CC'$ at point $O$. $OC'$ is perpendicular to $A'C'$ at $C'$.}
    \label{2-d_image}
\end{figure}

Write $l$ as the length of steel wire rope $OA$. Note that $OG$ is the gravity direction and thus is perpendicular to $GC$ and $GA$. Then we have $GA = l\sin\alpha$ and $OG=l\cos\alpha$ in the right triangle $\triangle OGA$. Next, considering the right triangle $\triangle OGC$, we have $OC=l\cos\alpha/\cos\beta$ and $GC=l\cos\alpha\tan\beta$. Then we apply the law of cosine in $\triangle AOC$ and $\triangle AGC$ respectively, and obtain $AC^2 = l^2\cos^2\alpha/\cos^2\beta+l^2-2l^2\cos\alpha\cos\angle AOC/\cos\beta$ in $\triangle AOC$ and $AC^2 = l^2\cos^2\alpha\tan^2\beta+l^2\sin^2\alpha-2l^2\cos\alpha\tan\beta\sin\alpha\cos\gamma$ in $\triangle AGC$, which leads to $\cos\angle AOC = \cos\alpha\cos\beta+\sin\alpha\sin\beta\cos\gamma$. Note that $\angle AOC$ and $\angle A'OC'$ are vertical angles, so the values of which are equal. In the right triangle $\triangle A'OC'$, $OC'$ is perpendicular to $A'C'$. Then, we have $\cos\angle A'OC' = h / \sqrt{m^2+h^2}$; see Figure \ref{2-d_image} for a more detailed graphical illustration. We then have
\begin{equation} \label{geo_formula}
    (m^2+h^2)(\cos\alpha\cos\beta+\sin\alpha\sin\beta\cos\gamma)^2=h^2.
\end{equation}
Consider an ideal case with $\beta=0$. This is the case when the video camera is perfectly installed so that the principle axis of the camera perfectly overlaps with the gravity direction. Then, \eqref{geo_formula} can be simplified as $(m^2+h^2)\cos^2\alpha = h^2$. This further suggests that $\cos\alpha = h/\sqrt{m^2+h^2}$.

\begin{figure}[!ht]
    \centering
    \includegraphics[width=0.35\textwidth]{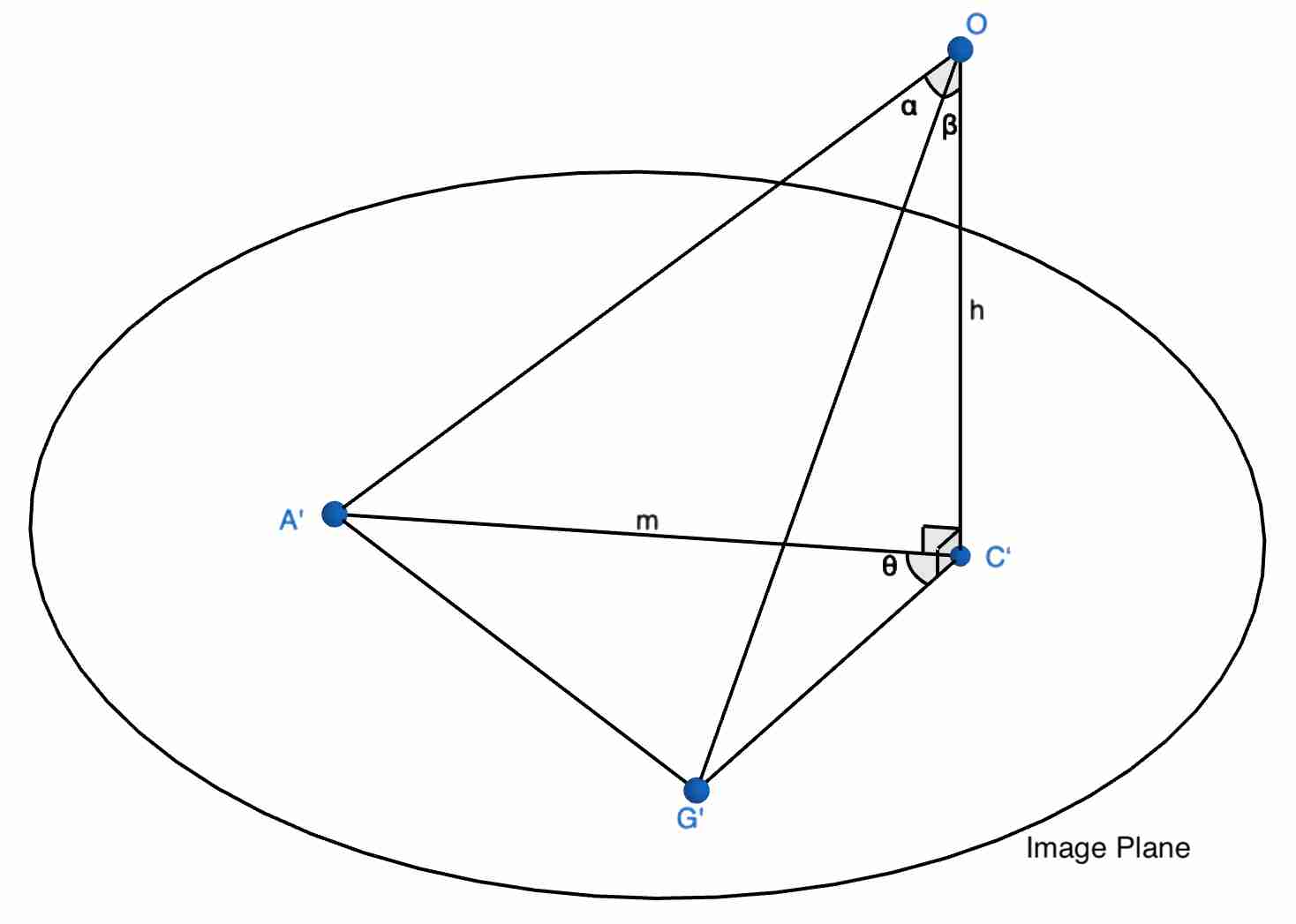}
    \caption{The graphical geometric illustration containing the camera $O$ the image plane $A'C'G'$. $OC'$ is perpendicular to plane $A'C'G'$ at $C'$.}
    \label{imageplane}
\end{figure}

Through equation \eqref{geo_formula}, we are able to estimate the value of $\beta$. However, we could not solve the value of $\alpha$ since the value of angle $\gamma$ is not observed. Write $\theta=\angle A'C'G'$. Note that $OC'$ is the principle axis of the camera and thus is perpendicular to both $C'G'$ and $C'A'$. Therefore, we should have $OG'=h/\cos\beta$ and $C'G'=h\tan\beta$ in $\triangle OC'G'$. Recall that in the right triangle $\triangle OC'A'$, $OA'=\sqrt{m^2+h^2}$. Apply the law of cosine, we have $A'G'^2=m^2+h^2+h^2/\cos^2\beta-2h\sqrt{m^2+h^2}\cos\alpha/\cos\beta$ in $\triangle A'OG'$ and $A'G'^2 = m^2+h^2\tan^2\beta-mh\tan\beta\cos\theta$ in $\triangle A'C'G'$ (see Figure \ref{imageplane}), then we could derive that
\begin{equation} \label{formula_pred_alpha}
   \cos\alpha = \big(m\sin\beta\cos\theta+h\cos\beta\big)/\sqrt{m^2+h^2}.
\end{equation}
\subsection{The focal length and the camera angle\label{focallength}}

By equation \eqref{formula_pred_alpha}, we know that the swing angle $\alpha$ can be estimated as long as the values of $m,\theta,h$ and $\beta$ can be obtained. The values of $m$ and $\theta$ can be directly measured on the image plane. Since $\gamma$ on the ground plane follows a uniform distribution, the median of $\gamma$ as well as $\theta$ should be $0$ simultaneously. Therefore, we could empirically find a line through the center of the image plane, so that the sample median of the intersection angles formed by this line and $C'A'$ is as close as possible to $0$. Then, the value of $\theta$ for each image can be estimated. Next, how to estimate the camera angle $\beta$ and focal length $h$ becomes the key issue.

We start with the focal length. The focal length of a camera should be the distance between the center position of a group of lenses and the image sensor when the subject is focused. This is a key parameter about the camera and should be well specified in the manual of the video camera. In our case, the digital camera installed at the fly-jib head is produced by Hikvision (\url{https://www.hikvision.com/en/}), and its focal length should be $h^*=4.8$ mm. Unfortunately, the focal length $h^*$ given in the manual directly is not exactly what we want. The focal length provided in the manual is measured in terms of millimeter (mm). However, to make use of the geometric model in Figure \ref{geo_illus}, we need to transfer this length into pixel distance on the image plane. In other words, we can treat the distance between two consecutive pixels on the image place as one basic distance unit. We use $d_0$ to denote the pixel length for $1$ mm. Then, the focal length we need is given by $h = h^*d_0$. Subsequently, we need to calculate $d_0$ accurately. Fortunately, we know the image plane is on a $1/2.8''$ type CMOS image sensor, from which we have computed the diagonal length of the CMOS image sensor is $6.43$ mm. With a $1920\times 1080$ pixel size of the video, we could calculate that $1$ mm in our geometric model means $d_0=342.67$ lengths in terms of pixel distance. Therefore, we have $h = h^*d_0 = 1644.82$.

We next consider how to estimate the camera angle $\beta$. We propose here a statistical estimation method. In this regard, appropriate probability distribution assumptions need to be made for random variables $\alpha$ and $\gamma$. Specifically, we assume for $\gamma$ a uniform distribution in $[0,2\pi)$. This seems to be a reasonable assumption since no direction should be particularly favored by the swing motion. Next, we consider how to make a sensible assumption for the swing angle $\alpha$. Intuitively, the swing angle $\alpha$ should be more likely to be of small values than larger values due to the gravity effect. Therefore, we assume for it an absolute normal distribution. In other words, we assume that $\alpha$ can be written as $\alpha=|Z|$ for some $Z$ following a normal distribution with mean $0$ and variance $\sigma^2$. Furthermore, we assume $\alpha$ and $\gamma$ are mutually independent of each other. We made this independent assumption for two reasons. Practically, this independent assumption seems to be a fairly reasonable assumption. Recall that $\alpha$ and $\gamma$ stand for the swing angle and the random swing position angle on the ground plane. Therefore, regardless of the rotating position of the bucket grab (i.e., $\gamma$ changes), the swing angle $\alpha$ could be of any possible value without any particular preference related to $\gamma$. Theoretically, this assumption is also necessarily needed for simplifying the theoretical treatment.  Otherwise, the derivation of equation \eqref{MomentEstimator_beta_popu} will not hold and then no statistical estimates can be obtained; see equation \eqref{star} in \ref{append}. The consequence is that no statistical estimates can be obtained. In contrast, with the help of this assumption, we could then estimate the camera angle $\beta$ and the unknown variance parameter $\sigma^2$.

To estimate $\beta$ and $\sigma^2$, we develop here a novel iterative algorithm. To motivate our method, we take expectation at both sides of equation \eqref{geo_formula}. That leads to the following equation:
\begin{align}\label{MomentEstimator_beta_popu}
   E(h/\sqrt{m^2+h^2}) = \cos\beta\exp{(-\sigma^2/2)},
\end{align}
where the technical verification can be found in \ref{append}. Then, the equations \eqref{formula_pred_alpha}, (\ref{MomentEstimator_beta_popu}) and the fact $E(\alpha^2) = \sigma^2$ inspire the following iterative estimation method. Specifically, we assume a total of $n$ independent observations are generated and indexed by $i$. For the $i$-th observation, we have random variables $\alpha_i, m_i$ and $\theta_i$. To estimate $\beta$, we start with assigning for $\beta$ an initial value as $\Hat{\beta}^{(0)}=0$. Next, we use $\Hat{\beta}^{(t)}$ and $\Hat{\sigma}^{2(t)}$ to represent the estimator obtained in the $t$-th step. Then the estimator in the next step can be obtained as
\begin{align}
    \Hat{\alpha}^{(t+1)}_i &= \arccos\left\{\left(m_i\sin\Hat{\beta}^{(t)}\cos\theta_i+h\cos\Hat{\beta}^{(t)}\right)\Big/\sqrt{m_i^2+h^2}\right\},\label{est-1}\\
    \Hat{\sigma}^{2(t+1)} &= n^{-1} \sum_{i=1}^n \left\{\Hat{\alpha}_i^{(t+1)}\right\}^2,\label{est-2}\\
    \Hat{\beta}^{(t+1)} &= \arccos\left\{n^{-1} \sum_{i=1}^n h\Big/\sqrt{m_i^2+h^2} \exp{\left(\Hat{\sigma}^{2(t+1)}/2\right)}\label{est-3}\right\},
\end{align}
where equation \eqref{est-1} should be executed for every $1\leq i\leq n$. As one can see, equation \eqref{est-1} is the empirical version of equation \eqref{formula_pred_alpha} but has the unknown parameter $\beta$ replaced by $\Hat{\beta}^{(t)}$. Equation \eqref{est-2} is a standard moment estimator for $\sigma^2$ due to the fact $E(\alpha^2)=\sigma^2$. Ideally, we should use $\alpha_i$ in equation \eqref{est-2}. However, they are not directly observed. Therefore, they are replaced by their estimates $\Hat{\alpha}_i^{(t+1)}$. Equation \eqref{est-3} is the moment estimation of $\beta$ due to (\ref{MomentEstimator_beta_popu}) with the unknown parameter $\sigma^2$ replaced by $\Hat{\sigma}^{2(t+1)}$. The algorithm should be iteratively executed till convergence. Specifically, the termination criterion of the iterative algorithm is given by $|\widehat{\beta}^{(t+1)}-\widehat{\beta}^{(t)}|<\epsilon$ and $|\widehat{\sigma}^{2(t+1)}-\widehat{\sigma}^{2(t)}|<\epsilon$, where $\epsilon = 10^{-6}$ is set to be a tiny constant. By the time of convergence, we obtain the final estimators $\Hat{\beta}$ and $\Hat{\sigma}^2$.

\subsection{A simulation study\label{simulation}
}

To demonstrate the finite sample performance of the resulting estimators $\Hat{\beta}$ and $\Hat{\sigma}^2$, we present here a simulation study. Specifically, we fix $h=1,600$, $\beta = 5^{\circ}$, and $\sigma = 2^{\circ}$. Those parameters are particularly selected to be fairly close to the real situation. Thereafter, $\alpha_i$s and $\gamma_i$s are randomly generated from the absolute normal distribution and uniform distribution respectively for every $1\leq i\leq n$, where $n$ denotes the sample size. The value of $\cos \angle AOC$ and $m_{i}$s are calculated according to the equations in Section \ref{geometricanalysis}. Moreover, the value of $\cos\theta$ is calculated by equation \eqref{formula_pred_alpha}.

\begin{figure*}[!ht]
 	\centering
 	\subfigure[The simulation result of $\Hat{\beta}$]{\includegraphics[width=0.45\textwidth]{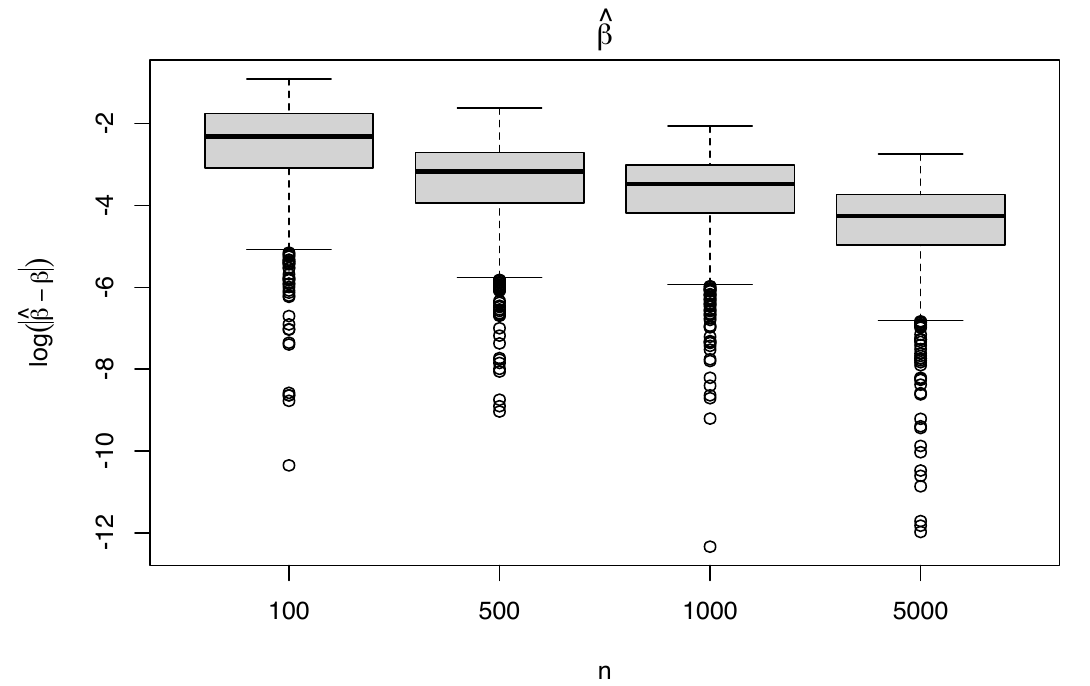}}
 	\hfil
 	\subfigure[The simulation result of $\Hat{\sigma}^2$]{\includegraphics[width=0.45\textwidth]{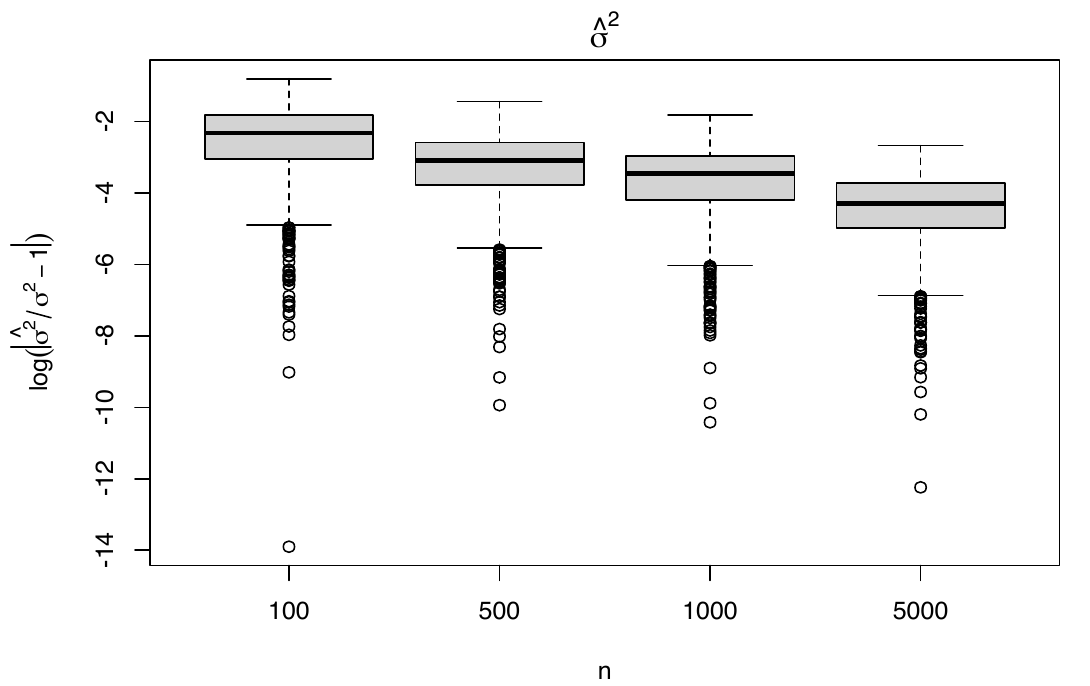}}
 	\caption{Log-transformed estimating errors for the proposed $\Hat{\beta}$ and $\Hat{\sigma}^2$ estimators.} 
 	\label{Simu_res}
\end{figure*}

Next, we apply the proposed iterative algorithm to compute $\Hat{\beta}$ and $\Hat{\sigma}^2$. Then the estimating errors are measured by $|\Hat{\beta} - \beta|$ and $|\Hat{\sigma}^2/\sigma^2-1|$ for $\hat{\beta}$ and $\Hat{\sigma}^2$ respectively. For a fixed $n$, the experiment is randomly replicated for a total of $M=1,000$ times. That leads to a total of $M$ values for the estimation error $|\Hat{\beta} - \beta|$ and $|\Hat{\sigma}^2/\sigma^2-1|$. They are then log-transformed and box-plotted in Figure \ref{Simu_res}.
By Figure \ref{Simu_res}, we find that the proposed algorithm is very accurate to estimate the unknown parameters. Specifically, as the sample size $n$ increases, the estimation errors for both $\Hat{\beta}$ and $\Hat{\sigma}^2$ steadily decrease towards $0$. Therefore, the statistical consistency of the proposed estimations is numerically confirmed.

\begin{figure*}[!ht]
 	\centering
 	\subfigure[The simulation result of computing $\Hat{\beta}^{(t)}$]{\includegraphics[width=0.45\textwidth]{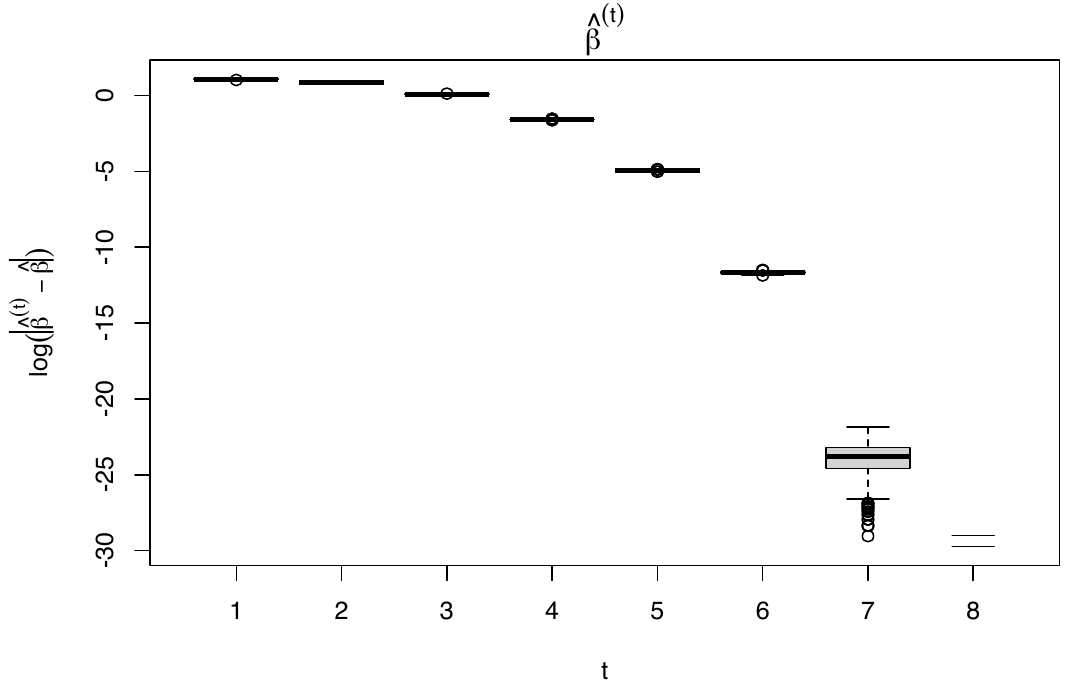}}
 	\hfil
 	\subfigure[The simulation result of computing $\Hat{\sigma}^{2(t)}$]{\includegraphics[width=0.45\textwidth]{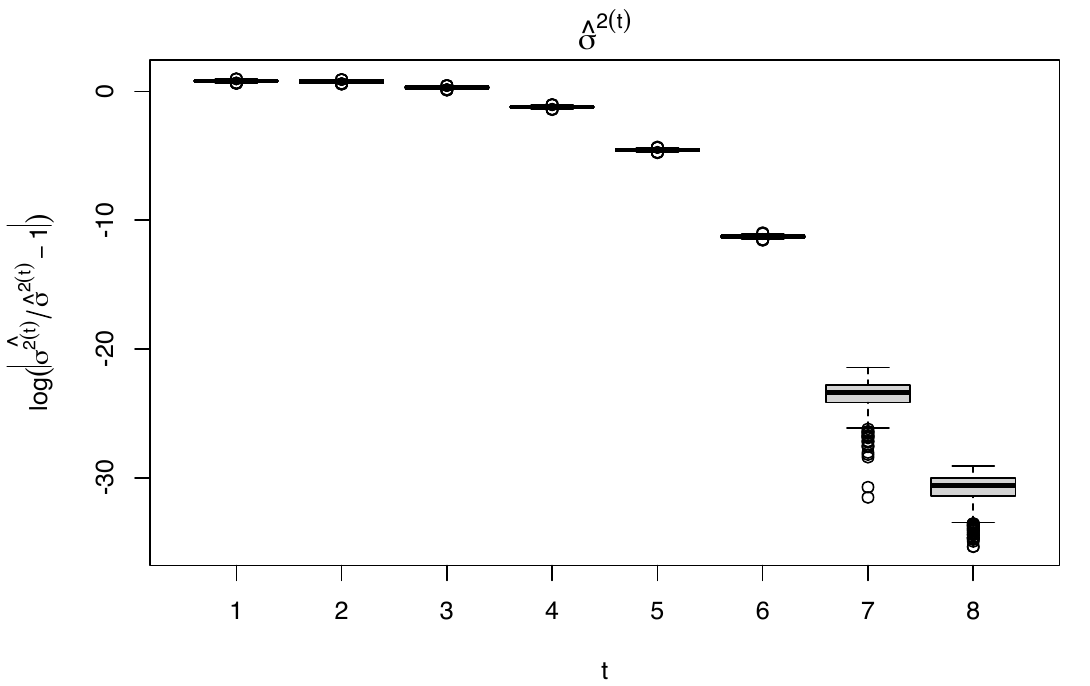}}
 	\caption{Log-transformed numerical errors for the proposed $\Hat{\beta}^{(t)}$ and $\Hat{\sigma}^{2(t)}$ estimators for different iteration steps.}
 	\label{Simu_comp}
\end{figure*}

Next, we want to demonstrate the numerical convergence properties of the proposed iterative algorithm. Recall that the experiment is randomly replicated for a total of $M=1,000$ times. For a given random replication $m$, we use $\Hat{\beta}_{(m)}^{(t)}$ and $\Hat{\sigma}_{(m)}^{2(t)}$ to denote the estimators $\Hat{\beta}^{(t)}$ and $\Hat{\sigma}^{2(t)}$ obtained in the $t$-th iteration of the $m$-th random replication. We should have $\Hat{\beta}_{(m)}^{(t)}\xrightarrow[]{}\Hat{\beta}_{(m)}$ and $\Hat{\sigma}_{(m)}^{2(t)}\xrightarrow[]{}\Hat{\sigma}_{(m)}^2$ as $t\xrightarrow[]{}\infty$, where $\Hat{\beta}_{(m)}$ and $\Hat{\sigma}_{(m)}^2$ stand for the final estimators $\Hat{\beta}$ and $\Hat{\sigma}^2$ obtained in the $m$-th random replication. Then, the numerical convergence rate of the proposed algorithm can be demonstrated by studying the numerical convergence errors given by $|\Hat{\beta}_{(m)}^{(t)} - \Hat{\beta}_{(m)}|$ and $|\Hat{\sigma}_{(m)}^{2(t)}/\Hat{\sigma}^2_{(m)}-1|$. For each fixed $t$, we have a total of $M$ numerical errors. They are then log-transformed and box-plotted in Figure \ref{Simu_comp}.
By Figure \ref{Simu_comp}, we find that both the two parameters of interest converge within a total of $8$ iteration steps for sufficient numerical accuracy. This implies that the proposed iterative algorithm converges at a very fast speed. Meanwhile, the CPU time consumed by each random replication is recorded. We find that the CPU time ranges from $0.026$s to $0.065$s with the median given by $0.028$s, which is fast enough for real time processing. Moreover, various initial values $\Hat{\beta}^{(0)}$ are tested. We find that the final estimator always converges to same numerical result.

\subsection{Empirical data analysis and results}\label{EDA}

We next apply the proposed method to the empirical data in hand. Recall that this is a video record lasting over $6$ minutes. It is collected from Qingdao Seaport (\url{https://www.qingdao-port.com/portal/en}). It contains a total of $n=3,699$ images with a $1920\times 1080$ pixel size. As mentioned before in Section \ref{datacp}, the detailed information about $m_i$ and $\theta_i$ is extracted from each image. For YOLOv5 implementation, various combinations of the optimizers are tested in this regard for parameter sensitivity. Specifically, different initial learning rates and final OneCycle learning rates are experimented for two popular optimizers, i.e., stochastic gradient descent \citep[SGD,][]{shalev2014understanding}, and Adam of \cite{kingma2014adam}. The best mAP@50:95 values on the validation dataset are recorded in Table \ref{Tab:exper-sensitivity}.
\begin{table}[!h]
\centering
\caption{Prediction Results of the YOLOv5 Model on the Validation Dataset.}
\label{Tab:exper-sensitivity}
\begin{tabular}{ccccccc}
\hline
\hline
\multirow{3}{*}{Initial Learning Rate} & \multicolumn{6}{c}{Final OneCycle Learning Rate}  \\
                                       & \multicolumn{3}{c}{SGD}  & \multicolumn{3}{c}{Adam} \\
                                       & 0.001   & 0.005  & 0.01  & 0.001   & 0.005  & 0.01  \\ \hline
0.0005                                 & 85.0      & 84.6   & 84.6  & 93.1    & 92.7   & 92.6  \\
0.001                                  & 86.4    & 87.2   & 87.0    & 93.8    & 93.5   & 93.4  \\
0.005                                  & 92.9    & 92.6   & 92.9  & 91.7    & 91.3   & 92.0   \\
0.01                                   & 93.8    & 94.5   & 94.2  & 90.1    & 90.5   & 90.5  \\ \hline
\end{tabular}
\end{table}
By Table \ref{Tab:exper-sensitivity}, we find that the optimizer of a combination for SGD with an initial learning rate of 0.01 and the final OneCycle learning rate of 0.005 performs the best. Therefore, we use this combination of learning rates for position extraction.

\begin{figure*}[!ht]
 	\centering
 	\subfigure[$m$]{{\includegraphics[width=0.3\textwidth]{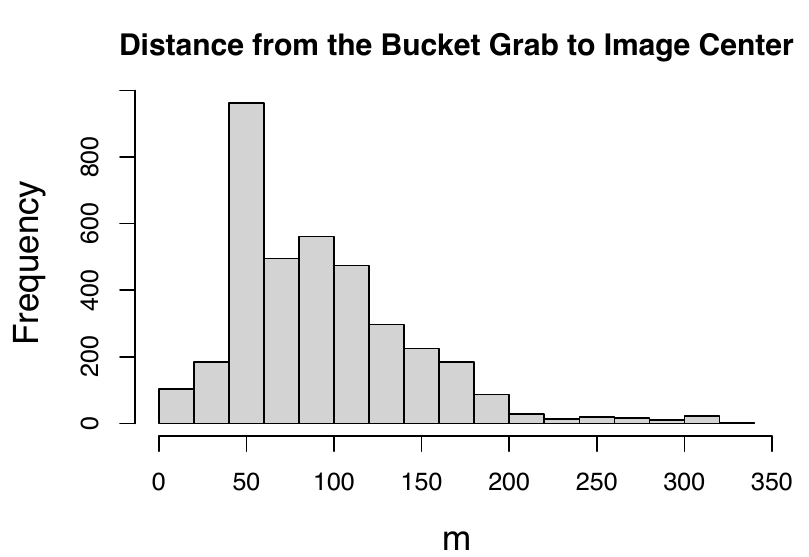}\label{desc-m}}}
 	\hfil
 	\subfigure[$\theta=\angle A'C'G'$]{\includegraphics[width=0.3\textwidth]{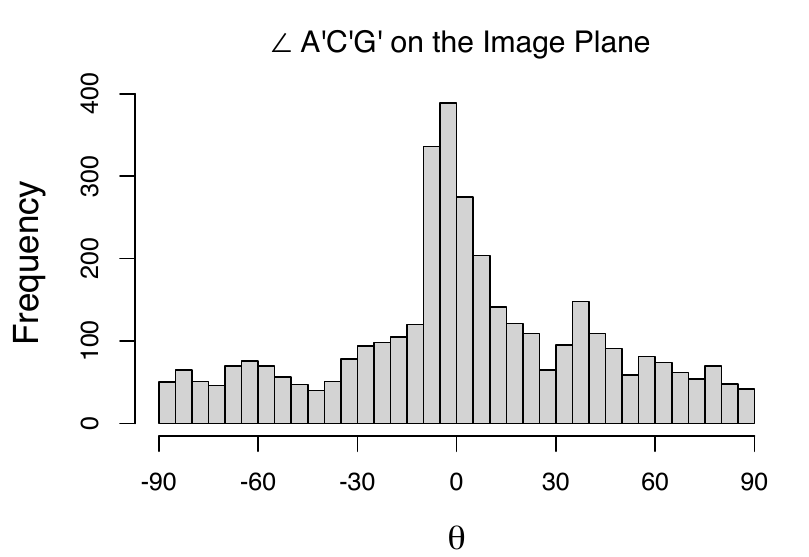}\label{desc-theta}}
 	\hfil
 	\subfigure[The Rope Length]{\includegraphics[width=0.3\textwidth]{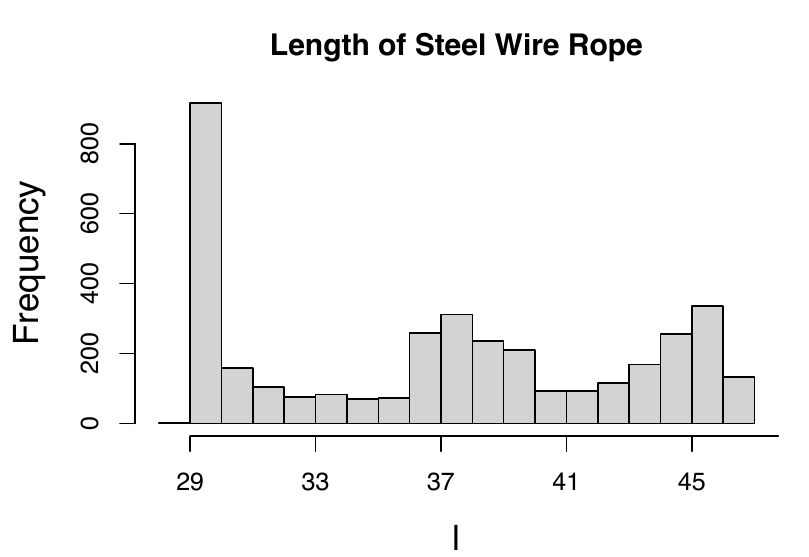}\label{desc-len}}
 	\caption{The numerical information extracted from the video dataset.}
 	\label{describe_stat}
\end{figure*}

For illustration purposes, the histograms of length $m_i$ and angle $\theta_i$ extracted from those images are plotted in the left and middle panels in Figure \ref{describe_stat}. In By Figure \ref{desc-m}, we find that the values of $m_i$s mainly concentrate with the interval $[28,182]$ with a $90\%$ coverage. A small proportion of the $m_i$ values is larger than $300$, which accounts for about $0.8\%$ of the total samples. By Figure \ref{desc-theta}, we find that the estimated $\theta$ values are fairly symmetric about $0$. For the sake of completeness, information on the length of steel wire rope is also extracted from video images and plotted in the right panel of Figure \ref{describe_stat}. By Figure \ref{desc-len}, we find that the length of the steel wire rope appears to be uniformly distributed, except when the rope is at its minimal length during the rotation of the bucket grab.

\begin{figure*}[!ht]
 	\centering
 	\subfigure[Histogram of estimated $\Hat{\alpha}_i$s]{{\includegraphics[width=0.4\textwidth]{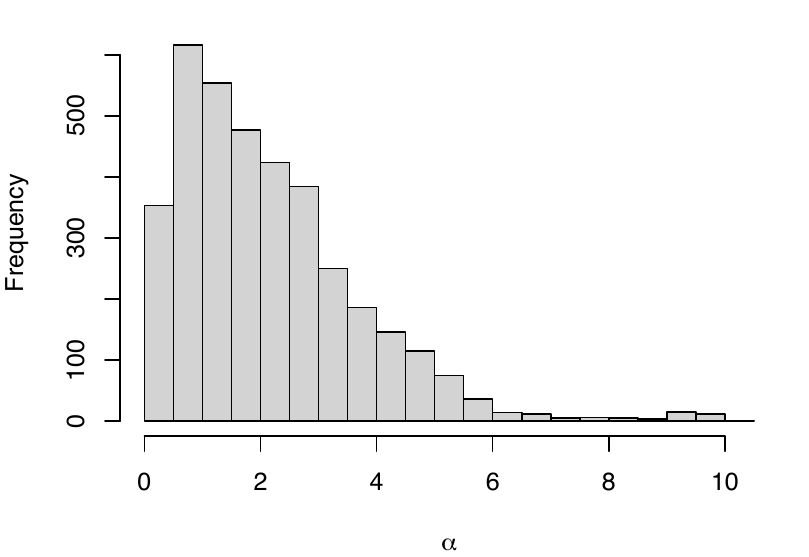}}}
 	\hfil
 	\subfigure[The frame with largest $\Hat{\alpha}$]{\includegraphics[width=0.5\textwidth]{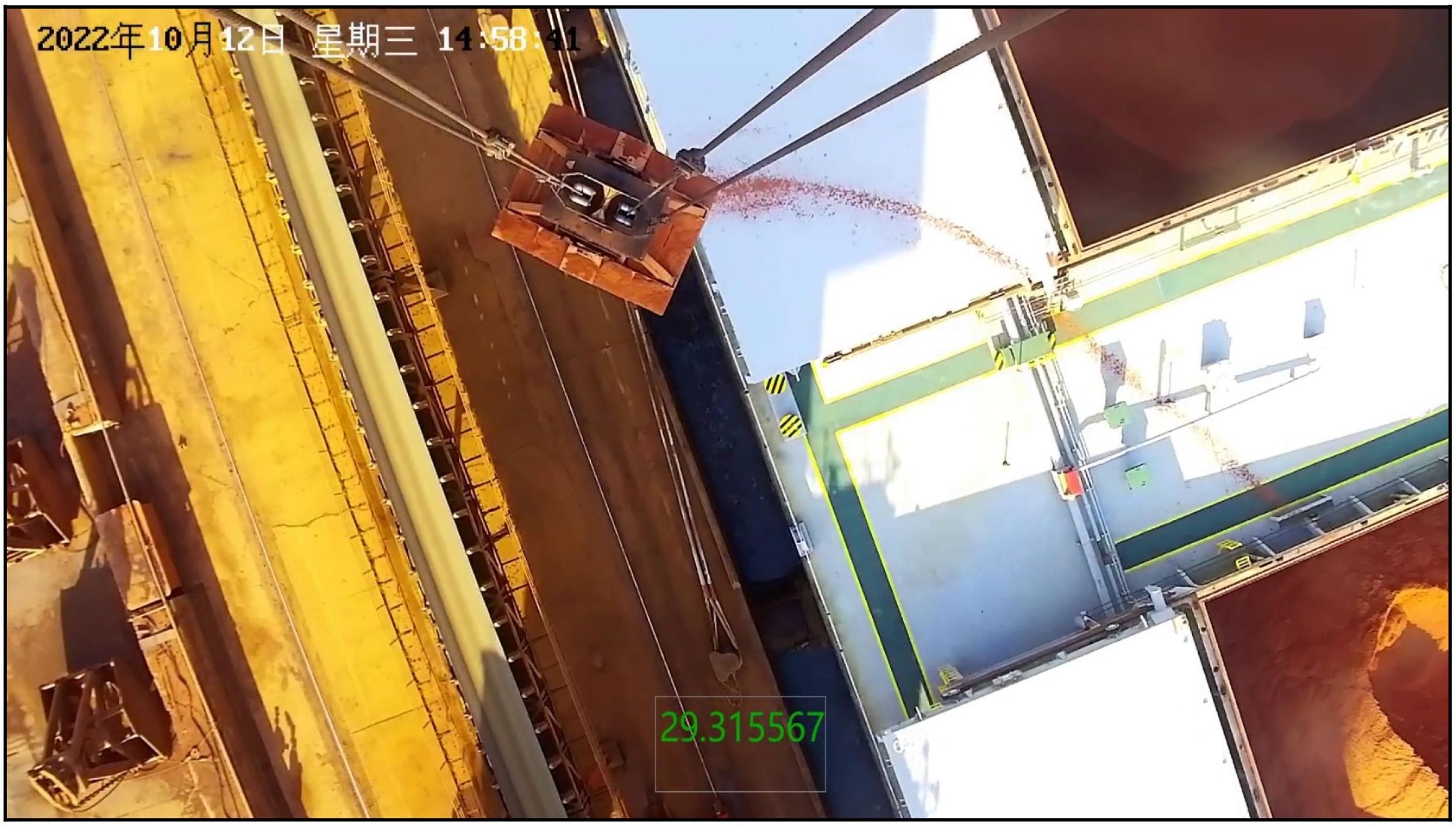}}
 	\caption{The left panel demonstrates the histogram of the estimated $\Hat{\alpha}_i$s, and the image with the largest estimated $\Hat{\alpha}$ is shown in the right panel.}
 	\label{Largest and small}
\end{figure*}

We then apply the iterative estimating algorithm (\ref{est-1}, \ref{est-2}, \ref{est-3}) to this dataset. The algorithm is executed for a total of $T=8$ iterations. The final numerical error measured by two consecutive estimators is given by $|\Hat{\beta}^{(T-1)}-\Hat{\beta}^{(T)}|=2.987\times 10^{-9}$ and $|\Hat{\sigma}^{2(T-1)}/\Hat{\sigma}^{2(T)}-1|=2.131\times 10^{-9}$. The final estimators are given by $\Hat{\beta}=2.530^{\circ}$ and $\Hat{\sigma} = 2.664^{\circ}$. This implies that the camera angle $\beta$ is clearly not $0$ even though small. The estimated value of $\Hat{\sigma}$ provides us a principal benchmark to construct a normal range for the swing angle. With the value of $\Hat{\beta}$, we are able to estimate for each image the swing angle $\alpha_i$. This leads to a total of $3,699$ values of $\Hat{\alpha}_i$ according to equation \eqref{est-1}. Their histogram is given in the left panel of Figure \ref{Largest and small}. By Figure \ref{Largest and small}, we find that the estimated swing angle is no larger than $5^{\circ}$ for most cases. This accounts for over $95\%$ of the total samples. This implies that in most cases the swing angle $\alpha_{i}$ is not large enough to pose a serious safety risk to the crane. However, we also notice that there do exist some cases with the estimated swing angle as large as $\Hat{\alpha}=10.010^{\circ}$. The image with the largest estimated swing angle $\Hat{\alpha}$ is provided in the right panel of Figure \ref{Largest and small}. It seems to us this is indeed a situation with an almost abnormally large swing angle. For the model training stage, the total CPU time consumed for training the model is $2.878$ second. Once the model parameters are estimated, they can be readily used to compute the swing angle for each image. For this stage, it takes about $0.009$ seconds to process one single image.

\subsection{Model validation}\label{validation}

It is remarkable that the proposed method is developed based on a number of assumptions. Specifically, the empirical accuracy of the proposed method relies on both the geometric model in Section \ref{geometricanalysis} and the statistical model in Section \ref{focallength}. Obviously, neither of them can be viewed as a perfect reflection of the real situation. They can only be regarded as approximations of reality. Therefore, we should reasonably expect that the true swing angle $\alpha_i$s should be different from the estimators $\Hat{\alpha}_i$s to some extent. However, how large is the difference? This is the most critical issue that calls for a method for validation.

\begin{figure}[!ht]
    \centering
    \includegraphics[width=0.75\textwidth]{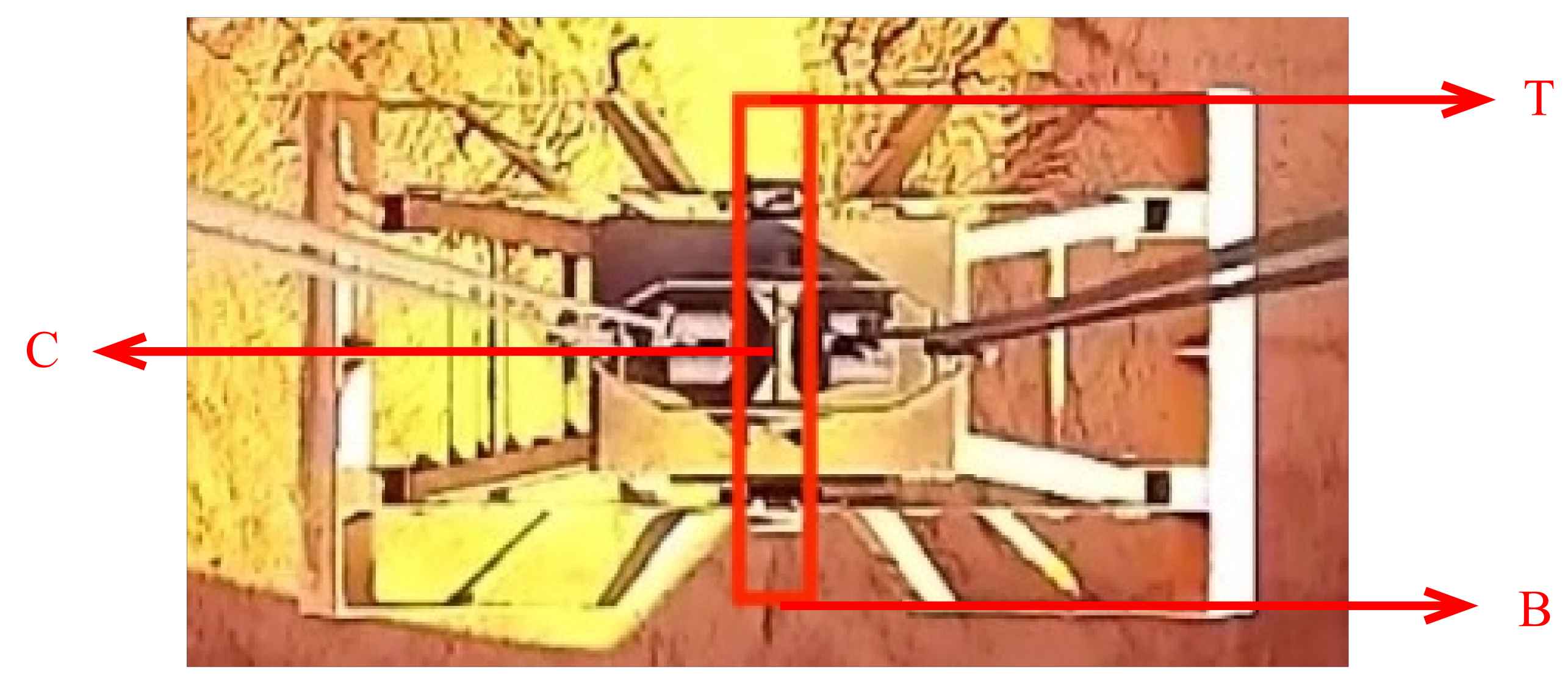}
    \caption{An enlarged bucket grab in the center of the figure.}
    \label{bucketgrab_width}
\end{figure}

\begin{figure}[!ht]
    \centering
    \includegraphics[width=0.5\textwidth]{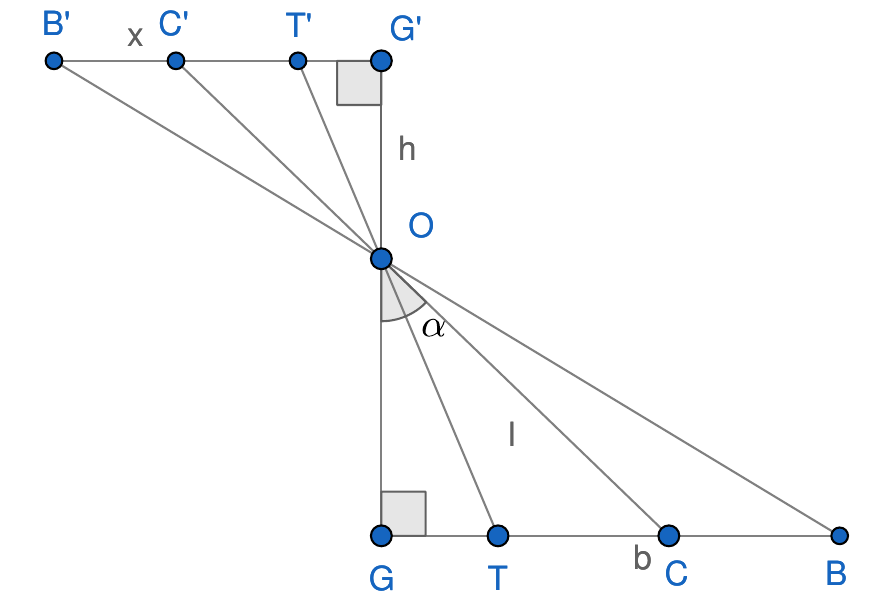}
    \caption{The illustration of the relationship considering the length of bucket grab in an ideal case with camera angle $\beta=0$.}
    \label{lengthgrab}
\end{figure}

To this end, we develop here a simple validation method as follows. We show in Figure \ref{bucketgrab_width} an enlarged figure about the bucket grab as captured by the video image, which is a figure manually selected. The width of the bucket grab is framed by the red box in Figure \ref{bucketgrab_width}. The top, bottom and central points of the grab width are represented by $T$, $B$, and $C$, respectively. The purpose here is to select an image with the bucket grab as close to the image center as possible. Furthermore, we wish the bucket grab is not arbitrarily rotated. Once this image is selected, the bucket grab width in terms of pixel distance, as shown by the red box in Figure \ref{bucketgrab_width}, can be measured manually. We find its width is about $141$ units in pixel distance.
Meanwhile, we know that the actual width of this bucket grab is $2.9$ meters. Then, the question is: can we provide another estimate for the swing angle by studying the geometric relationship between the width of the bucket grab in the ground plane in terms of meters and that in the image plane in terms of pixel distance? The answer is positive. A graphical illustration of this model is given in Figure \ref{lengthgrab}.
For simplicity, we assume an ideal case with $\beta=0$. Otherwise, the geometric relationship could be too complicated to be analytically tractable. Next, we use $T$ and $B$ to stand for the top and bottom endpoint of the bucket grab; see Figure \ref{bucketgrab_width} and \ref{lengthgrab}. We use $C$ to stand for the central point of the grab. Next, we use $B', T'$ and $C'$ to stand for their projections on the image plane. The pixel length of $B'T'$ can be observed on the image plane written as $x$. Since $OG$ is perpendicular to $GC$, we could simply derive that $OG=l\cos\alpha$ and $GC=l\sin\alpha$, then $GT=l\sin\alpha-b/2$ and $GB=l\sin\alpha+b/2$. Then, considering the similar right triangle $\triangle B'OG'$ and $\triangle BOG$, we have $xl\cos\alpha=hb$. Also, by $\tan\alpha = m/h$ on the image plane, we obtain $\sin\alpha=mb/xl$. Here we find that the focal length $h$ is not involved in the estimator $\Tilde{\alpha}=\arcsin\left(mb/xl\right)$ since the bucket grab width $b$ is used instead.

\begin{figure}[!ht]
    \centering
    \includegraphics[width=0.9\textwidth]{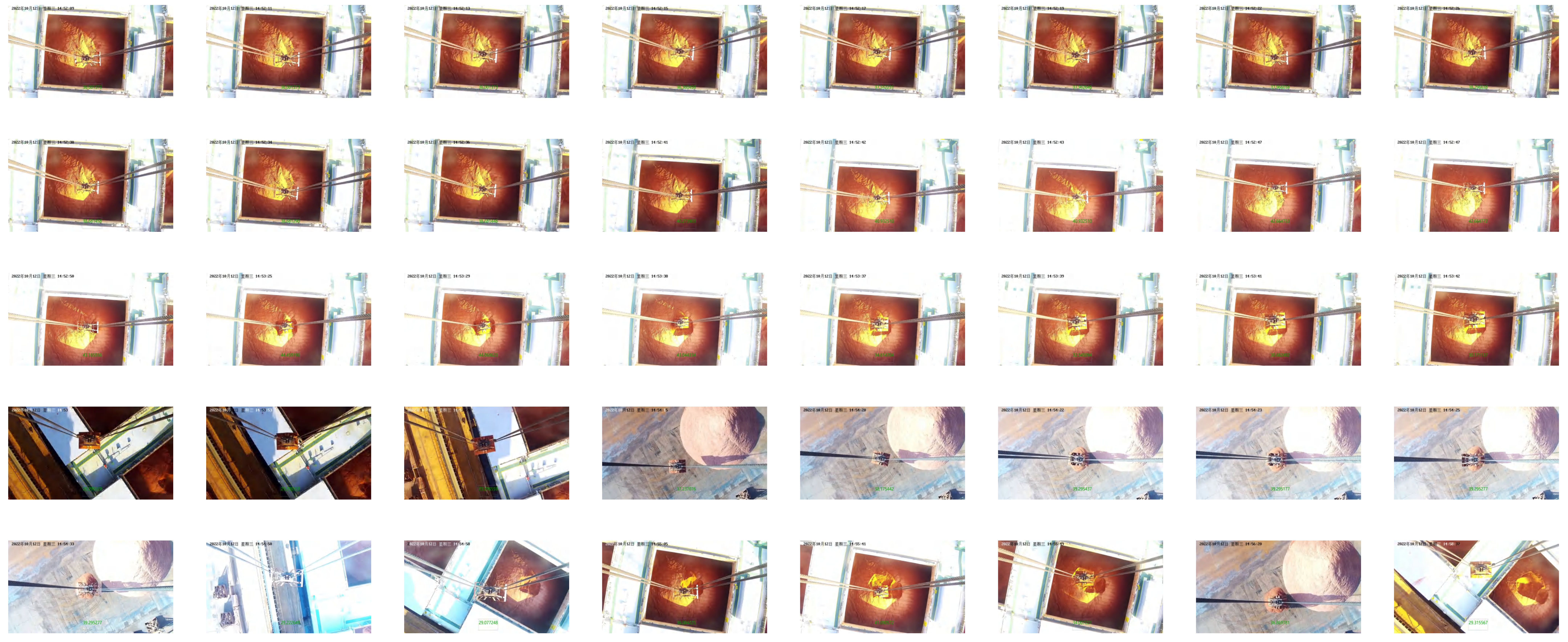}
    \caption{A snapshot of $40$ frame images without arbitrary rotations such that the image center is as possible as close to the straight line of the bucket grab width.}
    \label{40pics}
\end{figure}

\begin{figure*}[!ht]
 	\centering
 	\subfigure[The scatter plot of $\alpha$s from two models]{{\includegraphics[width=0.45\textwidth]{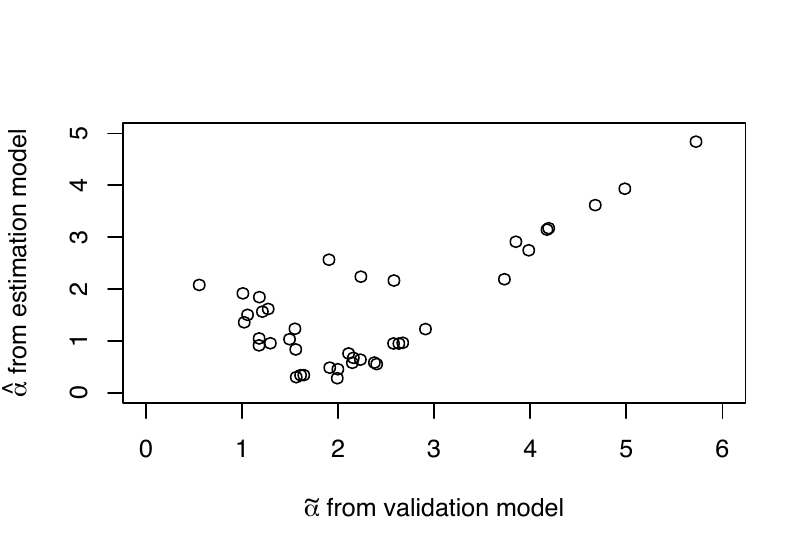}}}
 	\hfil
 	\subfigure[The boxplot of the difference $|\hat{\alpha}-\Tilde{\alpha}|$]{\includegraphics[width=0.45\textwidth]{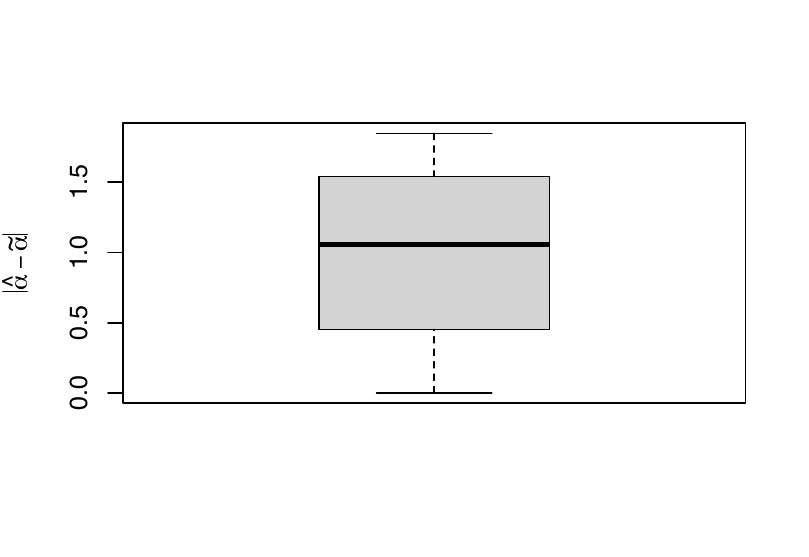}}
 	\caption{The comparison results of $\alpha$s between the estimation model and the validation model.}
 	\label{two-model}
\end{figure*}

Thereafter, another estimate for the swing angle for Figure \ref{bucketgrab_width} can be obtained, which is given by $\Tilde{\alpha}_i=1.337^{\circ}$. In contrast, the swing angle estimated by our iterative algorithm is $\Hat{\alpha}_i = 0.867^{\circ}$. The difference between $\Hat{\alpha}_i$ and $\Tilde{\alpha}_i$ is about $|\Hat{\alpha}_i-\Tilde{\alpha}_i|=0.470^{\circ}$, which seems a practically accepted difference. To further validate our model, we manually select a total of $40$ images with different estimated swing angles. A snapshot of these images is given in Figure \ref{40pics}.
We then compute both the validated swing angle $\Tilde{\alpha}_i$ and the estimated swing angle $\Hat{\alpha}_i$ for each image. Then the scatter plot can be obtained; see the left panel of Figure \ref{two-model}. We find that the two estimates tend to be more consistent with each other for situations with large swing angles. This happens to be the situation of the most practical importance. Moreover, the absolute differences $|\Tilde{\alpha}_i-\Hat{\alpha}_i|$ are then boxplotted in the right panel of Figure \ref{two-model}. We find that the largest difference is smaller than $2^{\circ}$, which seems practically very acceptable.

\subsection{Managerial implication} \label{Managerial implication}

As we have stated, the automated portal crane control system in Qingdao Seaport does not have any anti-swing ability. In order to avoid abnormal swings of the bucket grab, a field technician has to sit in front of the screens to monitor the entire operation process and manually control the bucket grab if an abnormal swing occurs. Obviously, the total number of cranes that can be monitored by one single technician is very limited. To solve this problem, the company running the seaport has to hire a large number of technicians to accomplish the task. This leads to a high cost in the labor force. Furthermore, the quality assurance across different technicians is another challenging issue. Consequently, we are motivated to develop here a computer vision based method for automatic swing angle detection.

\begin{figure}[!ht]
    \centering
    \includegraphics[width=0.92\linewidth]{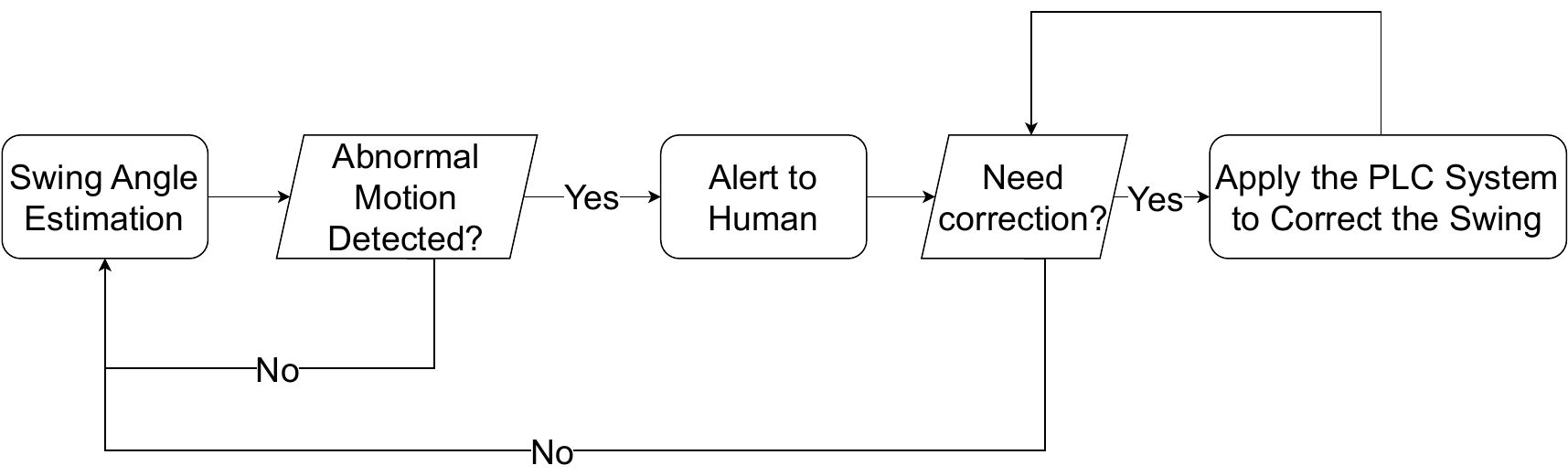}
    \caption{The flowchart of the managerial protocol of the seaport.}
    \label{flowchart-2}
\end{figure}

To practically implement our method, a standard protocol can be developed; see Figure \ref{flowchart-2} for a graphical illustration. Specifically, it contains a total of three steps. In the first step, our algorithm should be continuously executed for swing angle estimation. Once an abnormally large swing angle is detected (e.g., larger than $10^{\circ}$ according to the technical protocol at Qingdao Seaport), the system should immediately trigger an alarm for the staff. Then in the second step, the staff managing the crane should further confirm whether the situation is indeed dangerous or not. For the dangerous case, the staff should control the PLC system to drag the bucket grab back to its normal status. This anti-swing protocol is a standard operating procedure, which has been regularly used in the part. Our model helps to improve this protocol from a fully manual one to a partially automatic one, in the sense that the swing angle detection part becomes automatic but the anti-swing part remains to be manual.

\subsection{Comparison with existing methods} \label{Comparison with existing methods}

To conclude this section, we provide here a brief comparison between our algorithm and other methods in the literature. For a fair comparison, we only focus on those computer vision based methods.

\begin{figure*}[!ht]
 	\centering
 	\subfigure{\includegraphics[height=4cm]{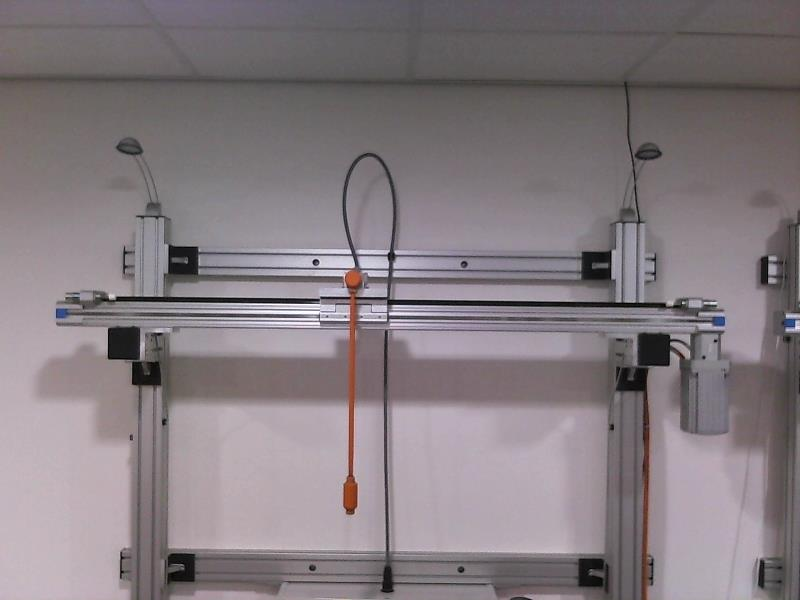}}
 	\hfil
 	\subfigure{\includegraphics[height=4cm]{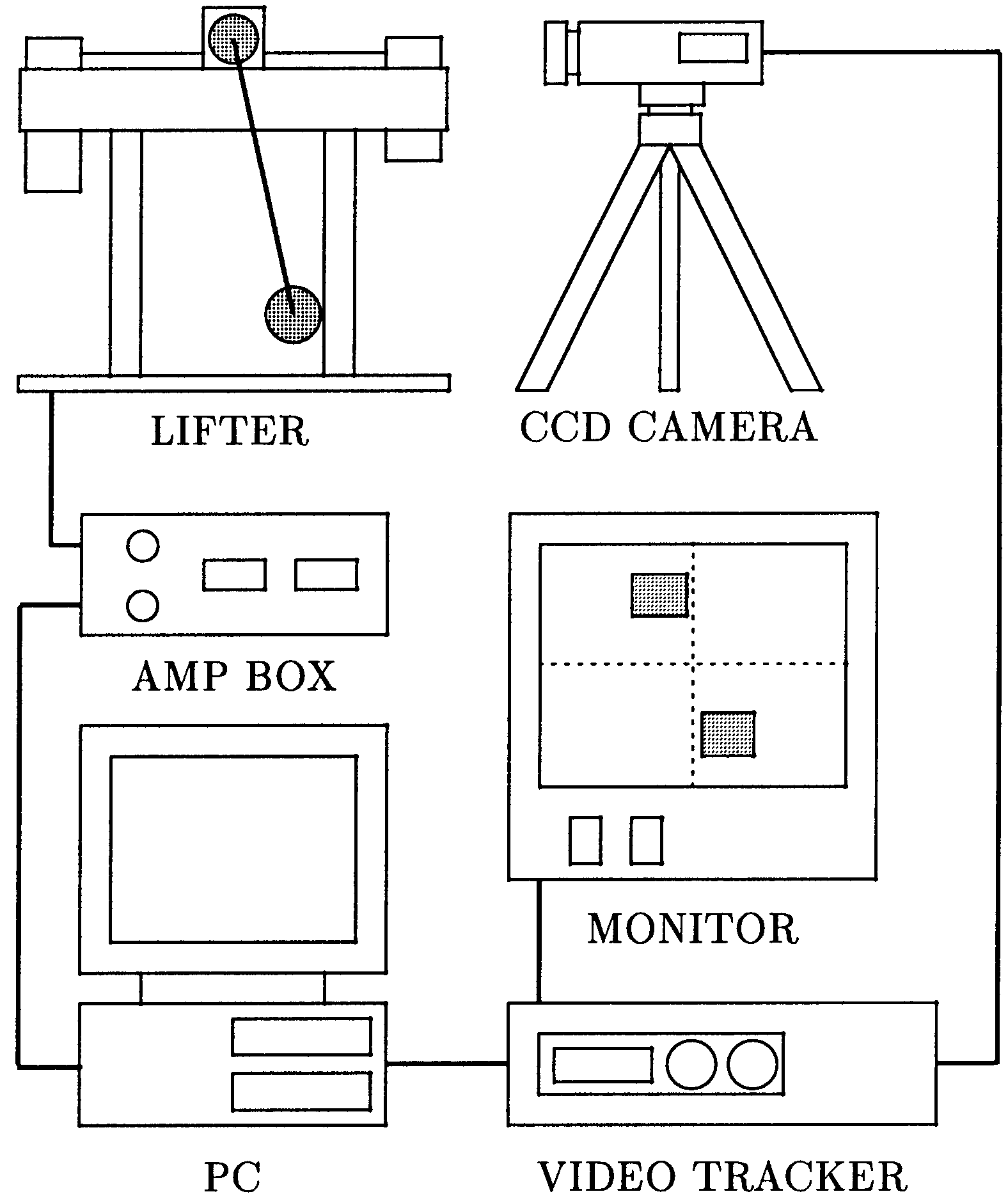}}
 	\caption{The methods of \cite{okubanjo2018vision} and \cite{matsuo2004nominal} for payload swing detection and anti-sway control with different camera positions.}
 	\label{comp-fig-1}
\end{figure*}

\begin{figure*}[!ht]
 	\centering
    \includegraphics[width=0.6\textwidth]{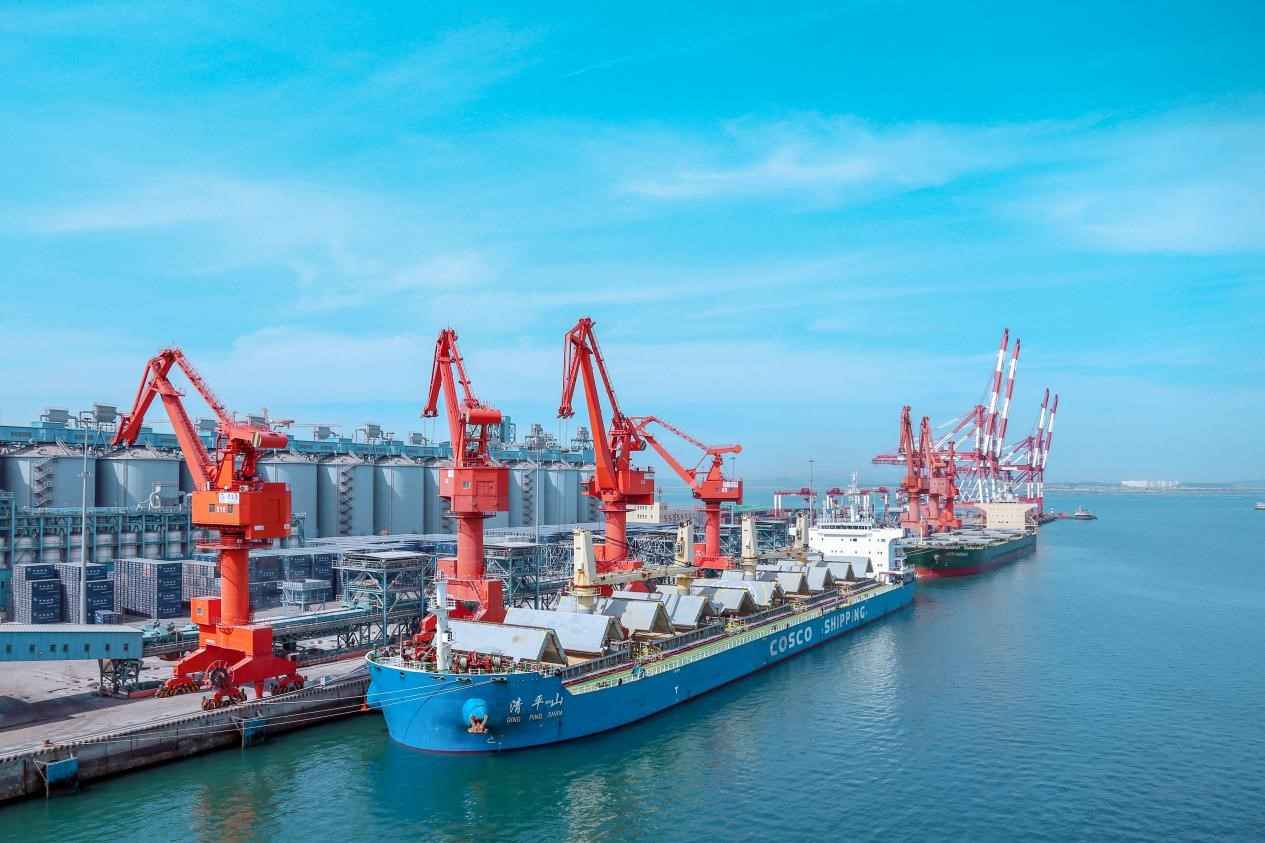}
 	\caption{A number of cranes working simultaneously at Qingdao Seaport.}
 	\label{cranes}
\end{figure*}

The first method to compare is the method of \cite{okubanjo2018vision}. This is a method based on the detection of the rope as well as the payload from video images. To implement this method, one needs to have a camera placed on the side of the overhead crane; see the left panel in Figure \ref{comp-fig-1} from Figure 1 of \cite{okubanjo2018vision} and the right panel in Figure \ref{comp-fig-1} from Figure 4 of \cite{matsuo2004nominal} for some illustration. By doing so, the swing motion can be fully captured. Unfortunately, their methods are tested only in laboratory environments like the left panel in Figure \ref{comp-fig-1}. Their practical performances remain unknown. Therefore, they cannot be used to solve our problem in Qingdao Seaport, since we do not have such a surveillance camera installed at the required position. Moreover, we suspect that placing the camera on the side of the overhead crane might not be a practically preferable choice. This is because there are too many cranes consecutively placed and operating simultaneously on the coast line at Qingdao Seaport; see Figure \ref{cranes} for some illustration from the official website of Qingdao Seaport (https://www.qingdao-port.com/portal/en). A camera placed on the side of the crane should inevitably capture noisy information due to other non-target cranes. That makes the subsequent image data analysis difficult and inaccurate. In contrast, our method only utilizes an existing camera installed on the fly-jib head. The image signals collected at this position are much cleaner and more informative; see Figure \ref{firstframe} for some illustration. 

\begin{figure*}[!ht]
 	\centering
    \includegraphics[width=0.6\textwidth]{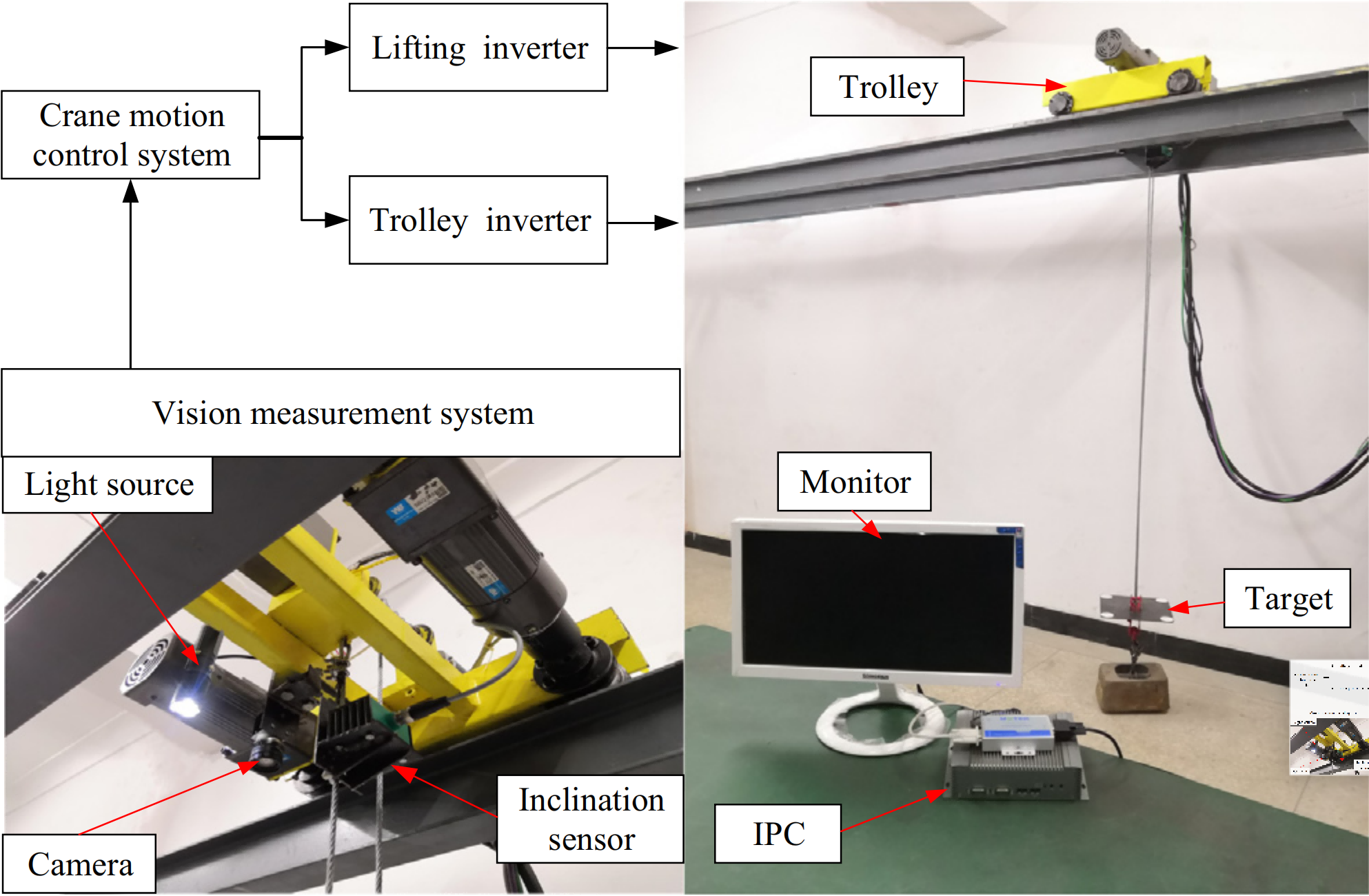}
 	\caption{The method of \cite{wu2020real} for payload swing detection and anti-sway control with markers painted on the payload for location.}
 	\label{comp-fig-2}
\end{figure*}

The second method to compare is the method of \cite{wu2020real}. This is a method based a simpler geometric model but without bias correction for the camera angle. It is remarkable that their overhead crane has a different physical structure from ours; see Figure \ref{comp-fig-2} from Figure 19 of \cite{wu2020real}. Furthermore, their method requires markers to be painted on the payload for easy detection. Unfortunately, we do not have markers painted on the bucket grabs in Qingdao Seaport for three reasons. First, the markers can be difficult to be detected if the grabs fall inside the bulk cargo, which is often the case for sea portal transportation. Second, the markers on the grab will inevitably be worn out with heavy-duty tasks. It is a time-consuming and costly task to fix the worn out markers by field technicians. Lastly, due to the safety regulations, the equipment and infrastructures at the port should be modified as little as possible.

To conclude this subsection, we compare our methods with the existing methods in Table \ref{Tab:comp}. Specifically, we focus on: (1) working environment request, (2) marker free or not, and (3) estimation accuracy.
\begin{table}[!h]
\centering
\caption{Comparison between the proposed method and existing methods in the literature.}
\label{Tab:comp}
\begin{tabular}{cccc}
\hline
\hline
Method              & \cite{okubanjo2018vision}    & \cite{wu2020real}   & Ours       \\ \hline
Working environment & Lab  & Lab & Seaport \\
Marker-free or not  & Yes  & No  & Yes     \\
Estimation accuracy & High & Relatively low & High   \\ \hline
\end{tabular}
\end{table}
As shown in Table \ref{Tab:comp}, we find that both the methods of \cite{okubanjo2018vision} and \cite{wu2020real} are not environment friendly. They can only be tested in a laboratory environment. Moreover, the method of \cite{okubanjo2018vision} requires placing the camera on the side of the crane, which is practically infeasible at Qingdao Seaport. The method of \cite{wu2020real} is marker-based, which cannot be implemented without painted markers. Additionally, the model does not account for potential camera angles in real seaport practice, which might lead to reduced estimation accuracy.

\section{Conclusion and future work}\label{conclude}

We develop in this work a novel spatial geometric model to describe the swing motion of a portal crane. This model takes the camera angle, camera focal length, and the random swing of the bucket grab into consideration. It can estimate the swing angle of an operating crane in real time. The most important feature of our method is the easiness of implementation. Our main contribution is the development of a computer-vision based marker-free method for bucket grab swing angle estimation. Compared with the existing methods, our method is: (1) working environment friendly, (2) marker free, and (3) highly accurate in swing angle estimation. Overall, the proposed method can be safely implemented to obtain an accurate and fast estimation of the swing angle of an operating crane. 

The main limitations of our method are as follows. First, since our method relies on a surveillance camera installed on the fly-jib head, it is difficult to adapt it to other applications with no cameras installed. Second, to validate the results produced by our algorithm, we require information about the steel wire rope length from the PLC system. For those applications without PLC systems, our method cannot be implemented directly.

To conclude this article, we present here a few directions for future work. First, it should be of great interest to test our model under various weather conditions. This is critically important for developing a practically robust algorithm. To partially address this issue, we have tested our model in two new experiments with snowy, windy, and dark conditions; see \ref{append-new} for a detailed discussion. Without any doubt, testing our model under even more diversified operation conditions is critically important. A significant amount of future research is inevitably needed. Second, it would be of great value if the bucket grab rotation angle can be estimated. This information is useful for integrating the shape of the bucket grab into our model. Third, it is remarkable that the image samples are extracted from a video of consecutive frames. There may exist a time series dependence for both the length of the steel wire rope and the swing angle. Taking the time series dependence into consideration might be helpful to estimate the swing angle with better accuracy. Fourth, it is of interest to establish the numerical convergence theories for our novel iterative estimating algorithm, even if both our simulation studies and real data analysis suggest that the resulting estimates are sufficiently accurate. Last but not least, it would be great if a golden standard about the swing angle can be practically developed.

\section*{Acknowledgements}

This work was supported by the National Natural Science Foundation of China [grant numbers 12271012, 11831008, 12101346]; the Open Research Fund of Key Laboratory of Advanced Theory and Application in Statistics and Data Science [grant numbers KLATASDS-MOE-ECNU-KLATASDS2101]; and the Shandong Provincial Natural Science Foundation [grant numbers ZR2021QA044].

\appendix 

\section{Technical derivations in Section 4.3}\label{append}

The equation \eqref{MomentEstimator_beta_popu} in Section 4.3 can be rigorously derived as follows
\begin{align}
    E(h/\sqrt{m^2+h^2})&=
    E(\cos\alpha\cos\beta+\sin\alpha\sin\beta\cos\gamma)\notag\\
    &= \cos\beta E(\cos\alpha) + \sin\beta E(\sin\alpha)E(\cos\gamma)\label{star}\\
    &= \cos\beta\int \sqrt{2/(\pi\sigma^2)}\cos x \exp{\left\{-x^2/(2\sigma^2)\right\}}{\rm d}x\notag\\
    &= \cos\beta\exp{(-\sigma^2/2)},\notag
\end{align}
where the second equality \eqref{star} holds since that $\alpha$ and $\gamma$ are assumed to be independent of each other.

\section{Two additional experiments}\label{append-new}

\begin{figure*}[!ht]
 	\centering
  \subfigure{\includegraphics[width=0.45\textwidth]{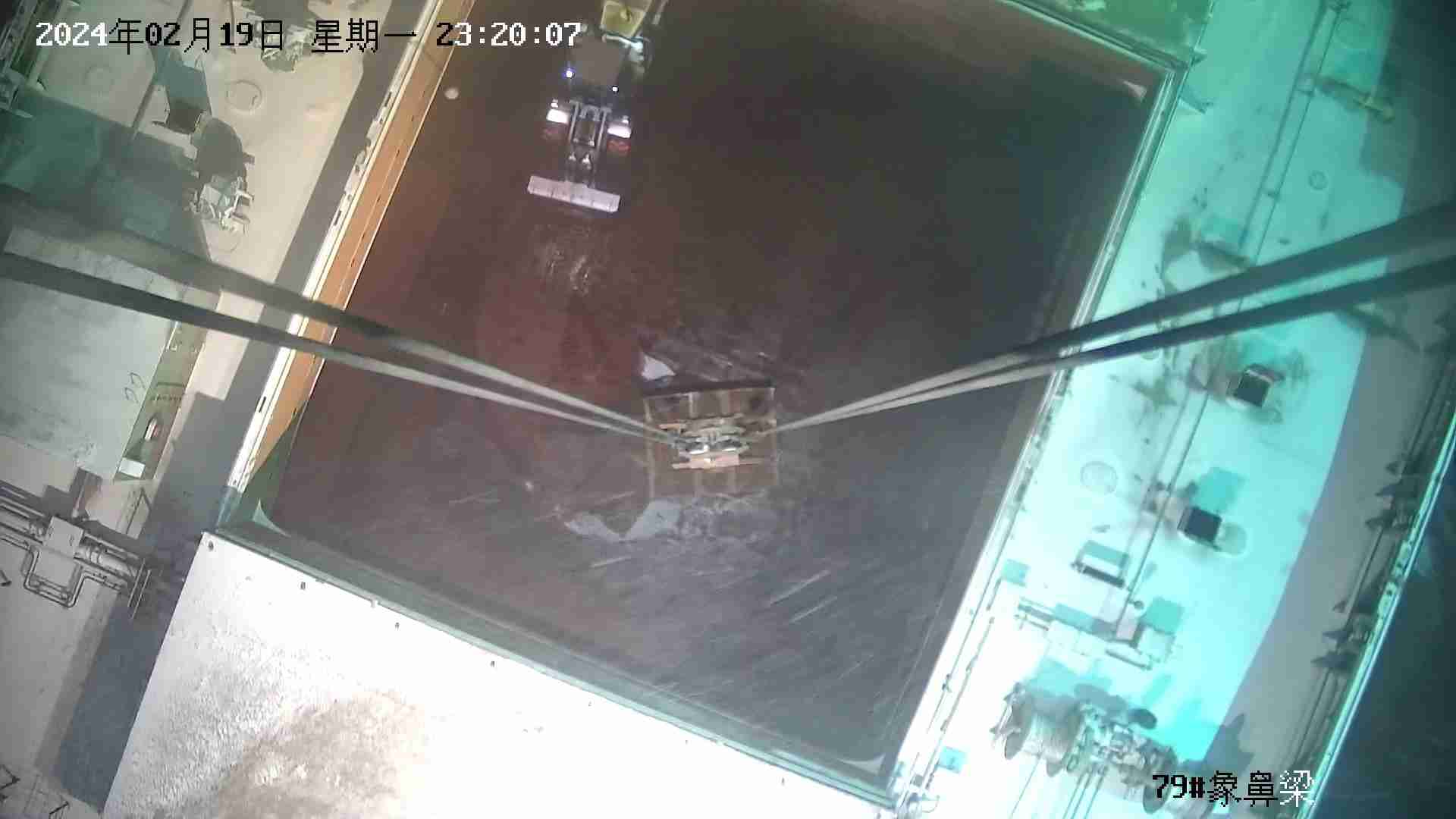}}
 	\hfil
 	\subfigure{\includegraphics[width=0.45\textwidth]{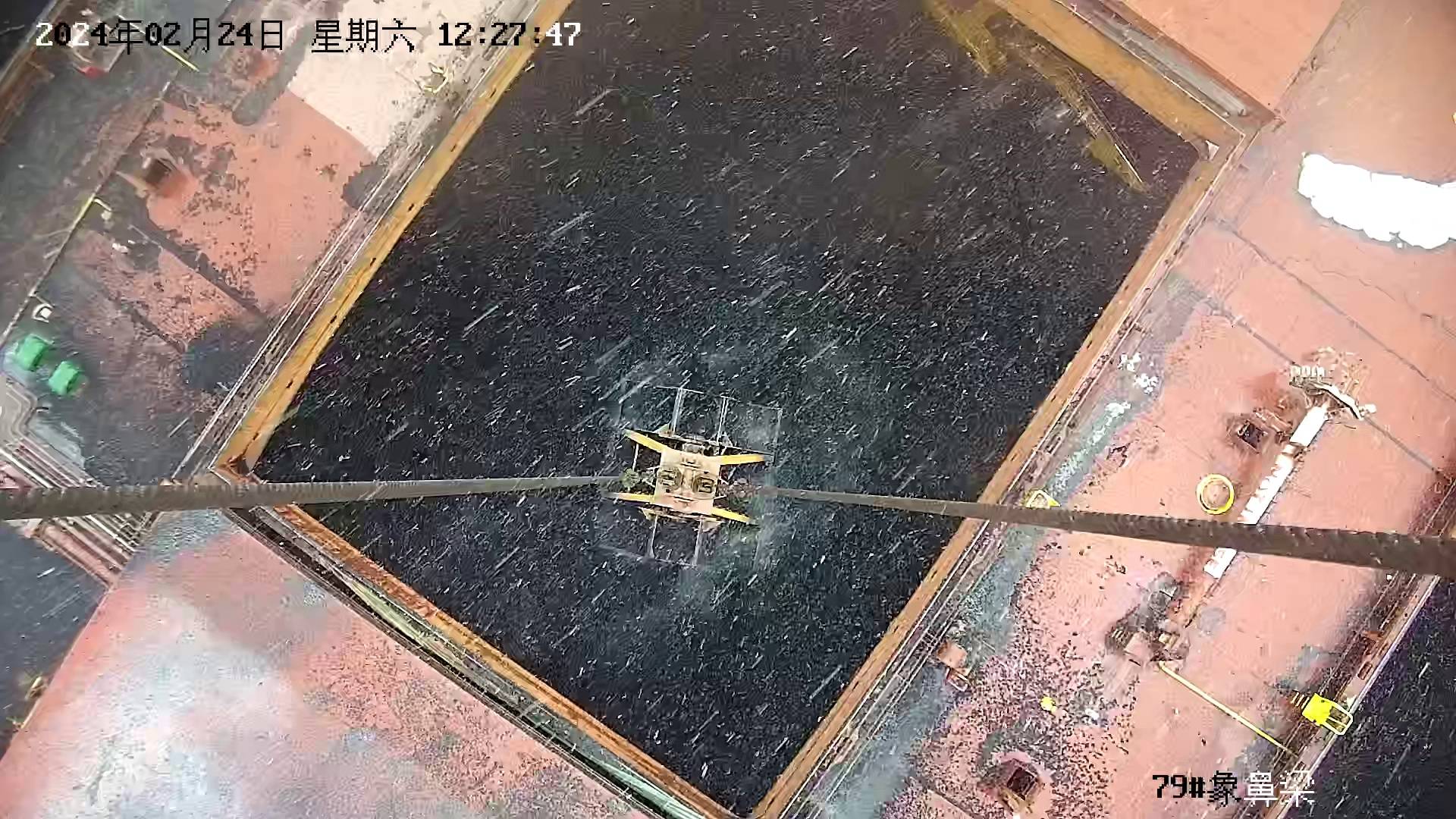}}
 	\caption{The left panel shows a video frame taken under a dark and light snowy condition. The right panel is taken under a heavy snowy and windy condition.}
 	\label{snow-new}
\end{figure*}

To test our method under various weather conditions, we collect two additional video records from Qingdao Seaport. The first video record lasts for about $2$ minutes and the second one for about $3$ minutes. The first record is taken at night under a light snowy condition. The second one is taken in the daytime but under a heavy snowy and windy condition; see the two panels of Figure \ref{snow-new} for some illustrations.
The two video records are then proceeded in the same way as Section \ref{EDA}. 

Specifically, we randomly extract a total of $220$ images from each of the two video records. Bounding boxes are then provided for each image in each dataset. Both datasets are then randomly split into a training dataset with $80\%$ samples and a testing dataset with $20\%$ samples. Subsequently, the YOLOv5 model can be fine-tuned on the training datasets with excellent performance on the testing datasets (i.e., mAP@50:95s of $0.950$ and $0.964$ for each video record). Next, we apply the iterative estimating algorithm (\ref{est-1}, \ref{est-2}, \ref{est-3}) to both datasets. The algorithm converges in a total of $T=7$ and $T=9$ iterations for each dataset, respectively. The final estimators are given by $\Hat{\beta}=3.005^{\circ}$ and $\Hat{\sigma} = 3.020^{\circ}$ for the first video record, and $\Hat{\beta}=4.221^{\circ}$ and $\Hat{\sigma} = 3.177^{\circ}$ for the second one. The results from the first video record are quite close to the estimator obtained in the previous video record as reported in Section \ref{EDA}. The estimated camera angle from the second experiment is much larger than the other two due to the heavy windy condition.

\bibliographystyle{elsarticle-harv}
\bibliography{ref1}

\end{document}